\PassOptionsToPackage{table}{xcolor}
\documentclass[preprint,12pt]{elsarticle}

\usepackage{framed} 

\usepackage{multicol} 

\usepackage{nomencl} 
\usepackage{siunitx}
\makenomenclature

\renewcommand*\nompreamble{\begin{multicols}{2}}

\renewcommand*\nompostamble{\end{multicols}}

\usepackage{amsmath}

\usepackage{svg}

\usepackage{tabularray}

\usepackage{longtable}
\usepackage{array}
\usepackage{booktabs}

\usepackage{mdframed}
\usepackage{url} 
\usepackage{url}

\usepackage{natbib}
\usepackage{url} 

\usepackage{enumitem}

\usepackage{subcaption}
\usepackage{graphicx}
\setlist{nolistsep}
\setlist[itemize]{leftmargin=*}

\usepackage[shortcuts,acronym]{glossaries}

\usepackage{svg}

\usepackage{etoolbox}
\renewcommand\nomgroup[1]{%
  \item[\bfseries
  \ifstrequal{#1}{A}{Acronyms}{%
  \ifstrequal{#1}{S}{Sets}{%
  \ifstrequal{#1}{P}{Parameters}{%
  \ifstrequal{#1}{V}{Variables}{}}}}%
]}

\usepackage{amssymb}

\usepackage{geometry}
\usepackage{xurl}
\usepackage{hyperref}
\usepackage[table]{xcolor}
\newcommand{\rev}[1]{\cellcolor{blue!18}#1}
\newcolumntype{P}[1]{>{\raggedright\arraybackslash}p{#1}}
\usepackage{soul}
\sethlcolor{blue!18}
\soulregister\cite7
\soulregister\ref7
\soulregister\SI7
\soulregister\mathrm7
\soulregister\textbf7
\newcommand{\revpara}[1]{\ifhmode\hl{#1}\else\indent\hl{#1}\fi}

\journal{Renewable and Sustainable Energy Reviews }

\begin{document}

\begin{frontmatter}


\title{Hydrogen in Practice: A Global Review of Demonstration Projects from Production to Integration}


\author[inst1]{Yiyina Teng}

\author[inst2]{Shahab Dehghan}

\affiliation[inst1]{organization={Department of Electrical Engineering, Yanshan University},
            city={Qinhuangdao},
            country={China}}

\affiliation[inst2]{organization={School of Engineering, Newcastle University},
            city={Newcastle upon Tyne},
            country={UK}}

\author[inst1]{Xiaoqiang Guo\corref{cor1}}
\cortext[cor1]{Corresponding author. Email:gxq@ysu.edu.cn}
\author[inst1]{Shiqi Zhang\corref{cor2}}
\cortext[cor2]{Corresponding author. Email:s.zhang@stumail.ysu.edu.cn}

\author[inst3]{Yanming Wan}
\affiliation[inst3]{organization={CHN Energy},
            city={Beijing},
            country={China}}

\author[inst5]{Sara Walker}

\affiliation[inst5]{organization={School of Chemical Engineering, University of Birmingham},
            city={Birmingham},
            country={UK}}
\author[inst2]{Vladimir Terzija}

\begin{abstract}
Hydrogen technology has the potential to address various challenges related to integrating low- and zero-carbon energy sources, as well as achieving carbon neutrality. It provides valuable solutions for transitioning to net-zero emission energy systems on a global scale. The implementation of hydrogen production and integration demonstration projects has received significant attention around the world and strong support from governments, highlighting the crucial role of such initiatives in the advancement of hydrogen energy. However, research in this area remains limited, particularly in terms of comprehensive and systematic reviews of industrial practices under real-world conditions. To bridge this gap, this article provides an in-depth review of hydrogen production, storage, and industrial applications, summarizing key features of thirteen exemplary hydrogen demonstration projects from around the world. The goal is to offer insights and guidance for the large-scale deployment of hydrogen energy while identifying potential technological pathways for the future.
\end{abstract}


\begin{footnotesize}
\begin{keyword}
Case Analysis \sep Demonstration Project \sep Hydrogen Production \sep Industrial Applications \sep Renewable Energy.
\end{keyword}
\end{footnotesize}

\end{frontmatter}



\nomenclature[A]{HPPO}{Hydrogen peroxide to propylene oxide.}
\nomenclature[A]{ALK}{Alkaline.}
\nomenclature[A]{PPC}{Poly Propylene Carbonate.}
\nomenclature[A]{ESO}{Electricity System Operator.}
\nomenclature[A]{GVA}{Gross Value Added.}
\nomenclature[A]{FCR}{Frequency Control Reserve.}
\nomenclature[A]{LFC}{Load Frequency Control.}
\nomenclature[A]{FAT}{Full Activation Time.}
\nomenclature[A]{IPA}{Intermountain Power Agency.}
\nomenclature[A]{EMS}{Energy Management System.}
\nomenclature[A]{R\&D}{Research and Development.}
\nomenclature[A]{P2P}{Power-to-Power.}
\nomenclature[A]{SDGs}{Sustainable Development Goals.}
\nomenclature[A]{MCFC}{Molten Carbonate Fuel Cell.}
\nomenclature[A]{NESO}{National Energy System Operator.}
\nomenclature[A]{PEM}{Proton Exchange Membrane.}
\nomenclature[A]{APG}{Austrian Power Grid.}
\nomenclature[A]{FRR}{Frequency Restoration Reserves.}
\nomenclature[A]{IPP}{Intermountain Power Project.}
\nomenclature[A]{AREH}{Asian Renewable Energy Hub.}
\nomenclature[A]{SCADA}{Supervisory Control and Data Acquisition.}
\nomenclature[A]{CVM}{Cloud Virtual Machine.}
\nomenclature[A]{SPE}{Solid Polymer Electrolyte.}
\nomenclature[A]{YSZ}{Yttria-Stabilized Zirconia.}
\nomenclature[A]{HC/LC}{High/Low Concentration .}

\newpage
\begin{footnotesize}
\printnomenclature
\end{footnotesize}


\vspace{-0.75cm}
\section{Introduction}\label{sec:intro}
As global warming intensifies, many countries have upgraded their energy policies in recent years to deal with climate change. Carbon peaking, carbon neutrality, and net-zero carbon energy transition have become the basic consensus of the international community \cite{olabi2021a}. Hydrogen energy has attracted promising attention from policymakers and stakeholders globally due to its valuable features for net-zero carbon energy transition, including low or zero carbon emission, flexibility, and efficiency \cite{kova2021a}. \revpara{By 2022, 26 countries and economies had developed national hydrogen strategies, reflecting the growing role of hydrogen in national energy-transition planning \cite{IEA_GlobalHydrogenReview2022,liu2022a}.} Hydrogen energy is expected to be essential in various areas and energy sectors, including electricity, heat, and transport \cite{liu2015a}. The advancement of hydrogen energy is seen as the key to countries with major economies to promote the transition to net-zero emission integrated energy systems. To illustrate this point, a concise summary of the latest hydrogen policy developments is provided for the following examples: Germany, the United Kingdom (UK), the United States (US), Japan, and China. 

Germany formulated its first “National Hydrogen Energy Strategy” \cite{Germany'sHydrogenStrategy} in June 2020, establishing a unified framework for the future of hydrogen energy integration, production, transportation, consumption, and innovation of hydrogen energy. Additionally, European countries are committed to developing a transnational hydrogen network to propel hydrogen production and its integration \cite{ji2021a,cardinale2023a}. This work will span multiple countries and transport hydrogen from where it is produced to where it is consumed \cite{cardinale2023a}. Regarding application projects, Germany is building hydrogen energy stations, hoping to produce hydrogen through electrolysis of water storage and distribute hydrogen to hydrogen fuel cell vehicles \cite{thomas2020a}. Furthermore, energy stations are located near transportation hubs and cities. This solution addresses the problem of renewable energy fluctuations, providing clean fuel and promoting the advancement of hydrogen fuel cell vehicles. \revpara{According to the updated UK hydrogen policy framework, the UK aims to develop up to 10 GW of low-carbon hydrogen production capacity by 2030, with at least half expected to come from electrolytic hydrogen \cite{UKhydrogenstrategy,policypaperUKhydrogenstrategy}.} Regarding infrastructure and supply chain development, the UK has committed to establishing novel frameworks for hydrogen transport and storage systems by 2025. The North Sea Transition Deal will leverage the industry’s existing capabilities, infrastructure, and the potential for private investment to develop emerging technologies, covering hydrogen production and Carbon Capture, Utilization, and Storage (CCUS) \cite{mclaughlin2023a}. The UK also has a comprehensive hydrogen integration strategy with different national centers for hydrogen integration and utilization, including the HI-ACT and UK-HyRES national hubs with over £20m of research funding \cite{newhubs}. The US advances the complete hydrogen energy industry chain as another clear example. \revpara{In June 2023, the United States released the ``U.S. National Clean Hydrogen Strategy and Roadmap'', which focuses on the effective use of clean hydrogen, reducing the cost of clean hydrogen, and developing regional hydrogen networks \cite{roadmap2002a}.} The roadmap stresses hydrogen energy's role in driving decarbonization, supporting economic growth, and boosting job opportunities. It indicates extensive potential for hydrogen production, distribution, storage, and application. The US likewise utilizes solar energy to generate green hydrogen. Converting solar energy into hydrogen for storage and application \cite{hassan2024a}. This use case integrates the clean energy from solar power with hydrogen storage capabilities, offering backup power, energy storage, and a hydrogen fuel supply. This makes it especially suitable for regions with abundant solar energy resources. \revpara{Japan launched the world's first national hydrogen strategy, the ``Basic Hydrogen Strategy'', in 2017 and revised it in 2023. With a focus on carbon neutrality and the development of a hydrogen society, Japan continues to emphasize hydrogen use in fuel cells, mobility, industry, and energy-system integration \cite{liu2022a}.} In addition, Japan has established a model for infrastructural energy interconnection application and strives to realize the interconnection between different infrastructural energy sources, including electric energy, hydrogen energy, and battery storage \cite{lee2014a}. This integration helps balance power production and consumption and strengthens the dependability and resilience of integrated energy systems under different disruptions and failures through diversifying energy sources.

China also emphasizes the complete hydrogen energy chain, covering production, storage, transport, and application \cite{tian2022a}. By breaking through crucial core technologies and enhancing the independent controllability of equipment, the aim is to promote the in-depth integrated development of the industrial and innovation chain. \revpara{China released the ``Medium- and Long-Term Plan for the Development of the Hydrogen Energy Industry (2021--2035)'' in 2022, establishing a national framework for hydrogen production, storage, transport, and utilization \cite{huang2024a}.} The industrial layout for hydrogen production and consumption will also be well-organized and well-established. All the developments and innovations mentioned above will strongly support achieving the carbon-peaking target in China. China harnesses renewable energy, including solar and wind, to generate hydrogen through water electrolysis \cite{li2020a}. It encourages distributed hydrogen energy systems to provide clean energy, actively promotes hydrogen fuel cell vehicles, and encourages automobile manufacturers to develop further. Additionally, China is developing hydrogen-integrated multi-energy microgrids, utilizing renewable energy with storage and electrolyzer to produce hydrogen and provide renewable energy to remote areas \cite{shao2023a}. Multi-energy microgrids can address the energy supply problem in isolated regions and further promote the application of renewable energy. In other words, hydrogen-integrated multi-energy microgrids can provide a more sustainable, resilient, and reliable energy supply for different types of remote consumers and, more importantly, reduce reliance on national grids.

\revpara{To provide a more analytical and compact policy background, Table \ref{tab: comparative_policy_background} compares the release or update year, key hydrogen-production targets, priority application sectors, infrastructure and storage priorities, and major funding or employment expectations of the selected national hydrogen strategies.}

\begin{table}[!ht]
\centering
\caption{Comparative policy background of selected national hydrogen strategies.}
\label{tab: comparative_policy_background}
\scriptsize
\setlength{\tabcolsep}{3pt}
\renewcommand{\arraystretch}{1.08}
\begin{tabular}{P{0.145\textwidth} P{0.115\textwidth} P{0.245\textwidth} P{0.215\textwidth} P{0.245\textwidth}}
\toprule
\rev{\textbf{Country / policy}} &
\rev{\textbf{Release / update}} &
\rev{\textbf{Hydrogen production or supply target}} &
\rev{\textbf{Priority application sectors}} &
\rev{\textbf{Infrastructure, storage, funding, or employment expectations}} \\
\midrule
\rev{Germany / National Hydrogen Strategy} &
\rev{2020; updated in 2023} &
\rev{At least 10 GW of domestic electrolyzer capacity for green hydrogen production by 2030.} &
\rev{Green hydrogen, industrial decarbonization, transport, heating, and energy-system integration.} &
\rev{Expansion of hydrogen networks, import corridors, refueling infrastructure, and funding programs for electrolysis and hydrogen-related IPCEI projects \cite{Germany'sHydrogenStrategy,cardinale2023a,thomas2020a}.} \\
\addlinespace[1pt]
\rev{UK / UK Hydrogen Strategy and updated policy framework} &
\rev{2021; ambition increased in 2022 and updated subsequently} &
\rev{Up to 10 GW of low-carbon hydrogen production capacity by 2030, with at least half expected to come from electrolytic hydrogen.} &
\rev{Industrial fuel switching, power generation, transport, heat, and natural-gas-network decarbonization.} &
\rev{Hydrogen transport and storage business models, CCUS-enabled and electrolytic hydrogen projects, national research hubs, and employment and clean-growth expectations \cite{UKhydrogenstrategy,policypaperUKhydrogenstrategy,newhubs}.} \\
\addlinespace[1pt]
\rev{USA / U.S. National Clean Hydrogen Strategy and Roadmap} &
\rev{Released in 2023} &
\rev{10 MMT/year of clean hydrogen by 2030, 20 MMT/year by 2040, and 50 MMT/year by 2050; the regional H2Hubs program is expected to contribute about 3 MMT/year by 2030.} &
\rev{Industrial decarbonization, heavy transport, power generation, long-duration storage, and synthetic fuels.} &
\rev{Regional clean hydrogen hubs connecting production, processing, delivery, storage, and end-use networks, with federal support for regional hydrogen infrastructure \cite{roadmap2002a,unknowp-m}.} \\
\addlinespace[1pt]
\rev{Japan / Basic Hydrogen Strategy} &
\rev{2017; revised in 2023} &
\rev{Hydrogen and ammonia supply expansion toward 3 Mt/year by 2030, 12 Mt/year by 2040, and 20 Mt/year by 2050.} &
\rev{Fuel cells, mobility, power generation, industry, and hydrogen/ammonia supply chains.} &
\rev{Development of international hydrogen and ammonia supply chains, refueling infrastructure, fuel-cell deployment, and domestic hydrogen-technology industries \cite{liu2022a,lee2014a}.} \\
\addlinespace[1pt]
\rev{China / Medium- and Long-Term Plan for the Development of the Hydrogen Energy Industry (2021--2035)} &
\rev{2022} &
\rev{100,000--200,000 t/year of renewable hydrogen production by 2025, together with about 50,000 fuel-cell vehicles by 2025.} &
\rev{Transport, energy storage, power generation, industry, green hydrogen, and hydrogen-integrated multi-energy systems.} &
\rev{Coordinated development of hydrogen production, storage, transport, and utilization, rollout of hydrogen refueling stations, and regional demonstration projects \cite{huang2024a,li2020a,shao2023a}.} \\
\bottomrule
\end{tabular}
\vspace{-0.3\baselineskip}
\end{table}

This growing global interest in hydrogen energy not only emphasizes its importance in policy frameworks but also highlights the need for comprehensive research into its practical applications, particularly in the context of industrial and on-site implementations. Previous research has been conducted on different characteristics of hydrogen-integrated energy systems, focusing on mathematical modeling and lab-scale experimentation. In \cite{van2023a}, various energy management systems for hydrogen-based microgrids are evaluated, including technologies, optimization methods, and simulation tools. In \cite{fonseca2019a}, a systematic review is conducted on different design challenges of distributed energy systems that incorporate hydrogen as an energy carrier. In \cite{maestre2021a}, recent developments in hydrogen-integrated renewable systems with large-scale energy storage systems are analyzed to address the intermittency of energy resources. In \cite{yue2021a}, various hydrogen energy applications in power systems, including re-electrifying, producing, and storing energy, are reviewed and analyzed. Furthermore, it explores different characteristics of fuel cells and water electrolyzers, supported by experimental data. In \cite{martin2020a}, a systematic review of hydrogen-integrated energy systems is conducted, discussing and explaining various applications, including transportation. The impact of user reactions on hydrogen-integrated energy systems is essential, highlighting the importance of how user responses are handled. In \cite{zhang2023a}, different US patents are reviewed to systematically review energy management strategies that utilize hydrogen storage to enhance renewable energy systems from a practical perspective.

\begin{figure}[!ht]
    \centering
    \includegraphics[width=0.75\textwidth]{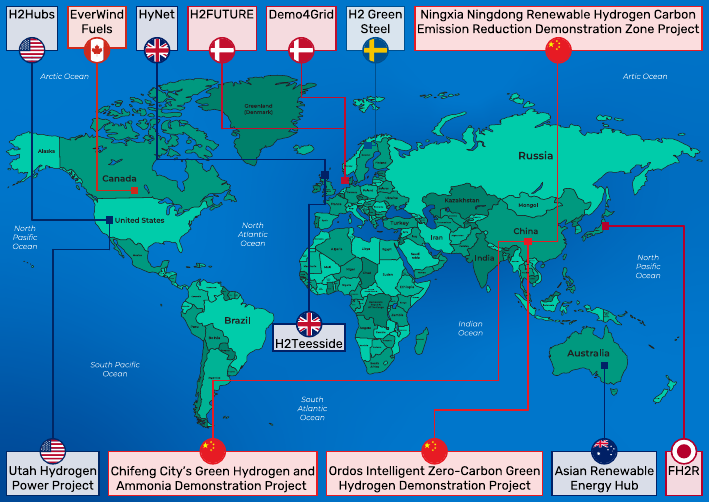}
    \caption{{Global distribution of hydrogen demonstration projects.}}
    \label{fig: Global distribution of hydrogen demonstration projects}
\end{figure}

In summary, previous research has reviewed different characteristics of hydrogen-integrated energy systems, including hydrogen production, storage, and transmission technologies, together with coordinated control, investment, and operation planning in integrated electricity and hydrogen networks \cite{aunedi2022a}. \revpara{Existing project-level studies have also mapped large portfolios of hydrogen and Power-to-X demonstration activities. For example, Chehade et al. reviewed 192 Power-to-X demonstration projects in 32 countries, while Wulf et al. identified 220 PtX research and demonstration projects in Europe by June 2020 \cite{chehade2019ptx,wulf2020ptx}. Therefore, the research gap should not be understood as an absolute absence of previous project-level reviews. Instead, the contribution of this work lies in its focused and comparative analysis of thirteen representative hydrogen demonstration projects across different regions, as shown in Figure \ref{fig: Global distribution of hydrogen demonstration projects}. Unlike broad project-mapping studies, this review emphasizes full-chain industrial integration, real-world operational constraints, cross-regional differences, and the coupling among hydrogen production, storage, transport, and end-use applications.} \revpara{This perspective provides complementary industrial-level insights for understanding the practical deployment of hydrogen-integrated energy systems and supports the future large-scale development of hydrogen energy.}

\section{Introduction to leading technologies in hydrogen demonstration projects}\label{sec: Introduction to leading technologies in hydrogen demonstration projects}

\subsection{Hydrogen Generation Technologies}\label{subsec: Hydrogen Generation Technologies}
In practical applications, various methods are available to produce hydrogen \cite{hermesmann2022a}. Table \ref{tab: Comparison of hydrogen production ways} outlines the comparative indicators and critical characteristics of typical types of hydrogen production. Gray hydrogen, generated from fossil fuels, is low cost, with prices typically between 0.7 and 2.3 \text{US\$/kg} \cite{incer-valverde2023a}. Although gray hydrogen is the dominant method worldwide, its reliance on fossil fuels results in considerable carbon emissions, thus impacting the environment. To address this problem, carbon capture technology is incorporated into blue hydrogen production. This process captures carbon dioxide generated during hydrogen production, reducing emissions released into the atmosphere. Although this reduces the amount of carbon dioxide released, it incurs additional costs, and thus, blue hydrogen costs more than gray hydrogen. Green hydrogen is produced by water electrolysis using renewable energy. In particular, advances in the green hydrogen sector could significantly accelerate global energy transitions toward net-zero emissions. Therefore, advances in the green hydrogen industry can significantly promote global net-zero emission energy transition developments. The key to these developments is price reduction since green hydrogen is much more costly than gray hydrogen, typically 2.4\raisebox{-0.6ex}[0pt][0pt]{\textasciitilde}7.1 \text{US\$/kg} \cite{zheng2012a}.

\begin{table}
\caption{{{Comparison of hydrogen production methods} \cite{hermesmann2022a}}.}
\resizebox{\columnwidth}{!}{%
\begin{tabular}{lllll}
\hline
Type           & Methods                          & Source           & $\mathrm{CO_2}$ emission& Cost   \\ \hline
Gray & Burning   fuels                  & Fossil   fuels   & High         & Low    \\
Blue & Burning   fuels + carbon capture & Fossil   fuels   & Medium       & Medium \\
Green & Electrolysis                     & Renewable energy & Low          & High   \\ \hline
\end{tabular}\label{tab: Comparison of hydrogen production ways}%
}
\end{table}

\revpara{Water electrolysis technologies used in hydrogen demonstration projects mainly include alkaline electrolysis (ALK), proton exchange membrane electrolysis (PEM), anion exchange membrane electrolysis (AEM), and solid oxide electrolysis cells (SOEC). However, these technologies differ substantially in commercial maturity and should not be presented as equivalent technology routes. Following recent U.S. DOE technology assessments, low-temperature alkaline electrolysis and PEM electrolysis are treated as commercial technologies, AEM electrolysis is closer to the pilot stage, oxygen-ion SOEC is at an early commercialization stage, and proton-conducting SOEC remains closer to the laboratory stage. Therefore, Table \ref{tab: Performance parameters of different water electrolysis technologies} distinguishes reported performance ranges from technology maturity, rather than presenting laboratory- or pilot-stage routes as equivalent to commercially mature technologies \cite{huang2023a,kumar2022a,DOEWaterElectrolysisAssessment2024}.}

\begin{table}[!ht]
\centering
\caption{Technical characteristics and maturity status of water electrolysis technologies \cite{ji2021a,hermesmann2021a,kumar2022a,DOEWaterElectrolysisAssessment2024}.}
\resizebox{\columnwidth}{!}{%
\begin{tabular}{lllll}
\toprule
\rowcolor{blue!18}
Comparison item & ALK & PEM & AEM & SOEC \\
\midrule
\rowcolor{blue!18}
Operating temperature (℃) & 60--90 \cite{ji2021a} & 50--80 \cite{kumar2022a} & 40--60 \cite{kumar2022a} & 700--850 \cite{kumar2022a} \\
\rowcolor{blue!18}
Electrolyte \cite{kumar2022a} & KOH/NaOH solution & Solid polymer membrane & \begin{tabular}[c]{@{}l@{}}Anion-exchange membrane /\\ alkaline medium\end{tabular} & YSZ or ceramic electrolyte \\
\rowcolor{blue!18}
Current density (A/\SI{}{\centi\metre\squared}) \cite{kumar2022a} & 0.2--0.8 & 1.0--2.0 & 0.2--2.0 & 0.3--1.0 \\
\rowcolor{blue!18}
$\mathrm{H_2}$ purity (\%) \cite{kumar2022a} & 99.5--99.9998 & 99.9--99.9999 & 99.9--99.9999 & 99.9 \\
\rowcolor{blue!18}
Pressure (bar) \cite{kumar2022a} & \textless{}30 & \textless{}70 & \textless{}35 & Around 1 \\
\rowcolor{blue!18}
Reported efficiency range (\%) \cite{kumar2022a} & 50--78 & 50--83 & 57--59 & \begin{tabular}[c]{@{}l@{}}Up to 89 under\\ high-temperature operation\end{tabular} \\
\rowcolor{blue!18}
Typical advantages & \begin{tabular}[c]{@{}l@{}}Mature technology;\\ relatively low cost\end{tabular} & \begin{tabular}[c]{@{}l@{}}High current density;\\ dynamic response\end{tabular} & \begin{tabular}[c]{@{}l@{}}Potential low-cost\\ catalyst route\end{tabular} & \begin{tabular}[c]{@{}l@{}}High efficiency with\\ heat integration potential\end{tabular} \\
\rowcolor{blue!18}
Main limitations & \begin{tabular}[c]{@{}l@{}}Lower current density;\\ alkaline corrosion issues\end{tabular} & \begin{tabular}[c]{@{}l@{}}Higher material and\\ catalyst cost\end{tabular} & \begin{tabular}[c]{@{}l@{}}Durability and scale-up\\ uncertainty\end{tabular} & \begin{tabular}[c]{@{}l@{}}High temperature, durability,\\ and integration challenges\end{tabular} \\
\rowcolor{blue!18}
Commercial maturity & \begin{tabular}[c]{@{}l@{}}Commercial\end{tabular} & \begin{tabular}[c]{@{}l@{}}Commercial\end{tabular} & \begin{tabular}[c]{@{}l@{}}Pilot-stage / emerging\end{tabular} & \begin{tabular}[c]{@{}l@{}}Oxygen-ion SOEC:\\ early commercial;\\ proton-conducting SOEC:\\ laboratory-stage\end{tabular} \\
\bottomrule
\end{tabular}\label{tab: Performance parameters of different water electrolysis technologies}%
}
\end{table}

\subsection{Different hydrogen storage technologies}\label{subsec: Different hydrogen storage technologies}
\revpara{To address the internal inconsistencies in the previous storage-technology comparison, this section has been reconstructed around measurable engineering indicators rather than only qualitative advantages and disadvantages. Table \ref{tab: Characteristics of different hydrogen storage technologies} now compares high-pressure gaseous storage, low-temperature liquid storage, and solid-material storage using indicators such as gravimetric capacity, volumetric density, operating pressure and temperature, boil-off or loss mechanisms, charging/discharging characteristics, safety concerns, cost level, and technology maturity.}

\begin{table}[!ht]
\centering
\caption{Measurable indicators of different hydrogen storage technologies \cite{zheng2012a,zuettel2004a,wang2023a,lim2010a,aziz2021a,DOEHydrogenStorage}.}
\small
\setlength{\tabcolsep}{3pt}
\renewcommand{\arraystretch}{1.18}
\resizebox{\columnwidth}{!}{%
\begin{tabular}{P{0.17\textwidth} P{0.27\textwidth} P{0.27\textwidth} P{0.27\textwidth}}
\toprule
\rowcolor{blue!18}
Indicator & High-pressure gaseous hydrogen storage & Low-temperature liquid hydrogen storage & Solid-material hydrogen storage \\
\midrule
\rowcolor{blue!18}
Storage principle & Hydrogen is compressed and stored as gas in high-pressure vessels. & Hydrogen is liquefied and stored under cryogenic conditions. & Hydrogen is stored through adsorption on surfaces or absorption/reaction within solid materials. \\
\rowcolor{blue!18}
Typical pressure & Usually high pressure; vehicle tanks commonly operate at 350--700 bar. & Generally stored at low pressure after liquefaction, but requires cryogenic containment. & Material-dependent; often low to moderate pressure, depending on the sorbent or hydride system. \\
\rowcolor{blue!18}
Typical temperature & Near ambient temperature. & Around $-253^{\circ}$C at atmospheric pressure. & Material-dependent; absorption/desorption often requires thermal management. \\
\rowcolor{blue!18}
Gravimetric storage capacity & Tank/system-level capacity is limited by pressure-vessel mass; reported ranges are typically around 1.0--5.7 wt\%. & Higher hydrogen mass fraction than compressed-gas systems in some applications, but system-level capacity is reduced by cryogenic insulation and auxiliary equipment. & Material-level capacity can be attractive, but usable system-level capacity is often reduced by host-material mass and heat-management requirements. \\
\rowcolor{blue!18}
Volumetric density & Lower than liquid hydrogen; limited by gas compressibility and tank pressure. & High volumetric density; liquid hydrogen density is about 70.8 kg/m$^3$ near the boiling point. & Potentially high volumetric density, but strongly dependent on material type and reversible usable capacity. \\
\rowcolor{blue!18}
Boil-off / loss mechanism & No cryogenic boil-off, but leakage and permeation must be controlled under high pressure. & Boil-off loss and liquefaction energy consumption are key constraints. & No cryogenic boil-off, but cycling losses may occur through incomplete absorption/desorption or material degradation. \\
\rowcolor{blue!18}
Charging/discharging characteristics & Fast refueling is possible, but thermal effects during fast filling and pressure management are important. & Transfer requires cryogenic handling; boil-off control is needed during storage and transport. & Absorption/desorption kinetics and heat transfer can limit charging and discharging rates. \\
\rowcolor{blue!18}
Main safety concerns & High-pressure rupture, leakage, and fire/explosion risks. & Cryogenic hazards, boil-off gas management, and insulation failure. & Material reactivity, impurity sensitivity, heat release/absorption, and long-term cycling stability. \\
\rowcolor{blue!18}
Cost and maturity & Commercially mature; widely used for vehicles, refueling stations, and stationary storage, but high-pressure tanks remain cost-intensive. & Commercial in specific sectors such as aerospace and special high-density storage applications; liquefaction and cryogenic systems increase cost. & Promising but less commercially mature for large-scale deployment; limited by material cost, usable capacity, kinetics, and durability. \\
\rowcolor{blue!18}
Typical applications & On-board storage, hydrogen refueling stations, tube trailers, and stationary buffer storage. & Aerospace, long-distance or high-density transport, and selected large-scale storage applications. & On-site buffering, niche stationary storage, and future mobility or marine applications where safety and volumetric density are prioritized. \\
\bottomrule
\end{tabular}\label{tab: Characteristics of different hydrogen storage technologies}%
}
\end{table}

\revpara{High-pressure gaseous hydrogen storage remains the most mature and widely deployed physical storage option. Hydrogen is compressed into durable tanks, with typical vehicle-storage pressures of 350--700 bar. This route benefits from relatively mature equipment, fast refueling capability, and broad applicability in fuel-cell vehicles, refueling stations, tube trailers, and stationary buffer storage. Its main limitations are lower volumetric density compared with liquid hydrogen, pressure-vessel cost and weight, and safety risks associated with high-pressure leakage or rupture.}

\revpara{Low-temperature liquid hydrogen storage provides higher volumetric density and is attractive when compact storage or long-distance transport is required. However, it requires cryogenic operation near $-253^{\circ}$C, and its overall performance is strongly affected by liquefaction energy consumption, insulation requirements, boil-off loss, and cryogenic safety management. Therefore, liquid hydrogen storage is commercially relevant in specific sectors such as aerospace and high-density transport, but it is not directly interchangeable with compressed gaseous storage for all demonstration-project conditions.}

\revpara{Solid-material hydrogen storage stores hydrogen through adsorption, absorption, or chemical reactions. It can offer advantages in safety and volumetric density because it may operate at lower pressures than compressed-gas tanks. However, its broader application is still constrained by material cost, reversible usable capacity, heat-management requirements, impurity sensitivity, absorption/desorption kinetics, and long-term cycling stability. Accordingly, it should be treated as a promising but less commercially mature route rather than as a reduced-cost and well-established technology.}

\revpara{Hydrogen storage vessels can still be classified into five types according to material composition, pressure capability, gravimetric capacity, maturity, and cost \cite{hassan2021a}. Type 1 and Type 2 vessels are mainly used for stationary or industrial storage, while Type 3 and Type 4 vessels are commonly applied in mobility scenarios such as fuel-cell vehicles and hydrogen refueling stations. Type 5 vessels remain under development and require further technical validation before large-scale deployment.}

\begin{table}[!ht]
\centering
\caption{Comparison of hydrogen storage vessels \cite{hassan2021a}.}
\resizebox{\columnwidth}{!}{%
\begin{tabular}{llllll}
\hline
Type   & Material                                   & \begin{tabular}[c]{@{}l@{}}Working\\ pressure\\ (MPa)\end{tabular} & \begin{tabular}[c]{@{}l@{}}Gravimetric\\ capacity or\\ tank mass(\%)\end{tabular} & \begin{tabular}[c]{@{}l@{}}Popularity\\ and maturity\end{tabular} & Cost   \\ \hline
Type 1 & Fully metallic                             & \textless{}50                                                      & -                                                                                 & ****                                                              & High   \\ \hline
Type 2 & Metallic enclosure and some fiber overwrap & Not limited                                                        & ±                                                                                 & **                                                                & Medium \\ \hline
Type 3 & Metallic liner and full composite overwrap & \textless{}45                                                      & +                                                                                 & *                                                                 & Low    \\ \hline
Type 4 & Full composite overwrap and polymer liner  & \textgreater{}100                                                  & +                                                                                 & *                                                                 & Low    \\ \hline
Type 5 & Fully composite                            & \multicolumn{4}{l}{Under consideration}                                                                                                                                                                                             \\ \hline
\end{tabular} \label{tab: Comparison of hydrogen storage vessels}%
}
\end{table}

\subsection{Different hydrogen utilization and applications}\label{subsec: Different hydrogen utilization and applications}
Power-to-Power (P2P) energy conversion systems can convert electrical energy into hydrogen energy using water electrolysis and convert hydrogen energy into electrical energy by fuel cells when necessary, constructing a closed-loop energy conversion system. This feature is essential to promote the substantial integration of renewable energy sources and implement the Sustainable Development Goals (SDGs) of the United Nations \cite{falcone2021a}. Fuel cell technology is a key part of a P2P energy conversion system, including Alkaline Fuel Cell (AFC), Proton Exchange Membrane Fuel Cell (PEMFC), Phosphoric Acid Fuel Cell (PAFC), Solid Oxide Fuel Cell (SOFC), and Molten Carbonate Fuel Cell (MCFC) \cite{zainal2024a,kumar2024a,dincer2015a}. \revpara{These fuel cell technologies differ substantially in maturity, operating temperature, fuel tolerance, start-up behavior, and efficiency definitions. Therefore, Table \ref{tab: fuel_cell_technologies} reports typical electrical efficiency separately from combined heat-and-power (CHP) performance where relevant, rather than using a single efficiency value for all application modes \cite{DOEFuelCellComparison,DOEMediumScaleCHP2011}.} The promising fast development of fuel cell technologies has reduced their investment costs \cite{staffell2019a}.

\begin{table}[!ht]
\centering
\caption{Comparison of different fuel cell technologies \cite{DOEFuelCellComparison,DOEMediumScaleCHP2011}.}
\resizebox{\columnwidth}{!}{%
\begin{tabular}{llllll}
\toprule
\rowcolor{blue!18}
Type & Electrolyte & Operating temperature & Typical electrical efficiency & Maturity / main use & Application areas \\
\midrule
\rowcolor{blue!18}
PEMFC & Polymer membrane & \begin{tabular}[c]{@{}l@{}}80--100℃, or higher\\ for HT-PEMFC\end{tabular} & Around 40--60\% & \begin{tabular}[c]{@{}l@{}}Commercially mature in\\ mobility and backup power\end{tabular} & \begin{tabular}[c]{@{}l@{}}Transportation and\\ distributed generation\end{tabular} \\
\rowcolor{blue!18}
AFC & \begin{tabular}[c]{@{}l@{}}Alkaline solution\\ or membrane\end{tabular} & Around 70℃ & Around 60\% & \begin{tabular}[c]{@{}l@{}}Commercial / niche\\ applications\end{tabular} & \begin{tabular}[c]{@{}l@{}}Backup power and\\ transportation\end{tabular} \\
\rowcolor{blue!18}
PAFC & Liquid phosphoric acid & Around 150--200℃ & \begin{tabular}[c]{@{}l@{}}Around 40\% for\\ electricity alone; above\\ 80\% only in CHP mode\end{tabular} & \begin{tabular}[c]{@{}l@{}}Commercial stationary\\ applications\end{tabular} & \begin{tabular}[c]{@{}l@{}}Distributed generation and\\ combined heat and power\end{tabular} \\
\rowcolor{blue!18}
MCFC & Molten carbonate & Around 600--700℃ & \begin{tabular}[c]{@{}l@{}}Around 45--60\%; higher\\ total efficiency in CHP mode\end{tabular} & \begin{tabular}[c]{@{}l@{}}Commercial stationary\\ applications\end{tabular} & \begin{tabular}[c]{@{}l@{}}Distributed generation and\\ combined heat and power\end{tabular} \\
\rowcolor{blue!18}
SOFC & Solid ceramic oxide & Around 600--1000℃ & \begin{tabular}[c]{@{}l@{}}Around 45--60\%; higher\\ total efficiency in CHP mode\end{tabular} & \begin{tabular}[c]{@{}l@{}}Commercial / demonstration\\ stationary applications\end{tabular} & \begin{tabular}[c]{@{}l@{}}Distributed generation and\\ combined heat and power\end{tabular} \\
\bottomrule
\end{tabular}\label{tab: fuel_cell_technologies}%
}
\end{table}

In addition to P2P, Power-to-Gas (P2G) energy conversion systems enable electrical energy to be converted to hydrogen by water electrolysis. Hydrogen can be introduced into existing natural gas pipelines \cite{ozturk2021a}. Blending hydrogen and natural gas can increase the low-carbon renewable content of natural gas networks. In addition, P2G can break down the barriers between electric power and natural gas systems. In other words, enabling cross-sector flexibility between integrated electric power and natural gas networks plays a key role in advancing multi-sector, low-carbon energy systems.

Moreover, Power-to-Gas + Gas-to-Power/Heat (P2P/H) systems enable hydrogen utilization in fuel cells and boilers to generate electricity and heat, therefore facilitating decarbonization of integrated electricity, heat, and transport systems as a clean alternative for fossil fuels \cite{aminudin2023a}. Similarly to P2P and P2G, P2P/H enables the linkage of multi-sector energy networks, providing additional cross-sector flexibility to manage the variability of renewable energy sources under different operational scenarios \cite{colbertaldo2018a,pastore2022a}. With the continuous progress of fuel cell technologies, hydrogen offers a wide range of potential as a fuel for multiple transportation types, such as cars, buses, trucks of various sizes, and airplanes \cite{yue2021a,aminudin2023b}. In this development process, hydrogen refueling stations have become critical terminal supply facilities supporting hydrogen fuel cell transportation. Based on different energy sources, refueling stations can be classified into: (i) in input-type refueling stations, hydrogen is supplied through a transportation network, with key components including compressors, storage containers, and fueling units, and (ii) self-sufficient refueling stations \cite{norouzi2021a}, including electrolysis equipment alongside the three core equipment mentioned above.

\subsection{Energy management system}\label{subsec: Energy management system}

A modern energy management system with advanced optimization models is essential to control hydrogen-integrated multi-sector energy systems, including different energy networks, consumers, producers, prosumers, and storage systems. With the substantial integration of renewable energy into integrated multi-sector energy systems, traditional energy management systems face additional challenges in controlling integrated multi-sector energy systems. In other words, considering spatial and temporal differences across various energy sectors, future energy management systems need further modernization to cost-effectively ensure the real-time, reliable, and resilient balance of production and consumption in integrated multi-sector energy systems with intermittent renewable resources \cite{taylor2022a}. In addition to modernizing energy management systems, transforming energy management regulations is necessary. For example, the Electricity System Operator (ESO) of the UK’s National Grid was transformed into the National Energy System Operator (NESO) in 2024. This transformation adopts a holistic approach to strategically planning and developing energy markets across different energy sectors to further support the UK government's climate target \cite{fso,acctrans}. 

Modeling multi-sector integrated energy systems involves complex interactions of multiple energy forms, making the development of advanced optimization tools for energy management systems a complex task \cite{dehghan2024a}. Firstly, holistic modeling of producers and consumers as well as transmission and distribution networks across different energy sectors is necessary, given the heterogeneity of energy \cite{kang2020a}. Therefore, mathematical models in energy management systems need to be designed with higher computational tractability in solving large-scale optimization problems in multi-sector integrated energy systems. Secondly, as integrated energy systems typically include distributed energy resources and energy storage devices, mathematical models must incorporate these components' dynamic characteristics \cite{mavromatidis2019a}. This requires accurate forecast and control mechanisms for intermittent sources such as solar and wind, along with precise modeling of the charging and discharging processes of energy storage devices \cite{he2023a}.

Due to potential techno-economic risks in utilizing hydrogen energy to promote the transition to future net-zero emission energy systems, it is necessary to exploit risk-aware energy management tools and models with modern control strategies to ensure the secure operation of multi-sector energy systems incorporating hydrogen under different techno-economic limitations cost-effectively \cite{yue2021a, liu2022b, vivas2018a, li2022a}. 

\section{Demonstration projects for hydrogen production in different countries across the world}\label{sec: Demonstration projects for hydrogen production in different countries across the world}

\subsection{Case selection and comparative evaluation framework}\label{subsec: Case selection and comparative evaluation framework}
\revpara{To strengthen the systematic nature of the project-level review, the thirteen cases were selected using four criteria: geographical diversity, coverage of different hydrogen pathways, availability of project-level public information, and relevance to industrial-scale deployment. The selected cases cover Europe, North America, Asia, and Australia and include low-carbon hydrogen production, renewable-electrolysis projects, hydrogen-based metallurgy, hydrogen-ammonia production, grid-interactive electrolysis, storage-based power generation, and regional hydrogen hub development.}

\revpara{The data used for project comparison follow a source hierarchy. Priority was given to official project documents, government or agency reports, and project-owner disclosures. Peer-reviewed literature and institutional reports were then used to support technical interpretation. Secondary media sources were used only when official or institutional data were unavailable or when recent project-status changes had not yet been reflected in formal project documents. Capacity and investment claims are therefore reported with their source year or status category wherever possible.}

\revpara{For consistency, the reviewed cases are classified as operational/completed demonstrations, under-construction projects, planned or announced projects, or policy/regional hub programs. This classification avoids treating verified operational projects, announced mega-projects, and funding programs as the same analytical category. H2Hubs is therefore described as a regional hub and infrastructure program rather than as a single hydrogen-production project; AREH is treated as a planned or prospective renewable-hydrogen and ammonia hub with updated status information; and H2 Green Steel is reported as Stegra, reflecting its updated project name and planned production timeline.}

\revpara{Because the public sources report heterogeneous indicators, the comparison table separates original reported values from annual hydrogen-output information. Annual hydrogen output is shown only when it is explicitly disclosed by the project source or can be reported without introducing hidden utilization assumptions. Hourly capacity, daily production, electrolyzer power, storage capacity, ammonia output, steel output, and regional hub targets are therefore retained in their original units and clearly identified as design capacities, announced targets, or program-level outputs rather than verified annual operational production. This treatment avoids the previous mixing of annual, hourly, daily, and power-capacity units and separates verified data from projected benefits.}

\subsection{European demonstration projects for hydrogen production}\label{subsec: European demonstration projects for hydrogen production}

\subsubsection{HyNet: UK’s low-carbon hydrogen production projects - Case 1}\label{subsubsec: HyNet: UK’s low-carbon hydrogen production project}

\begin{figure}[!ht]
    \centering
    \includegraphics[width=0.75\textwidth]{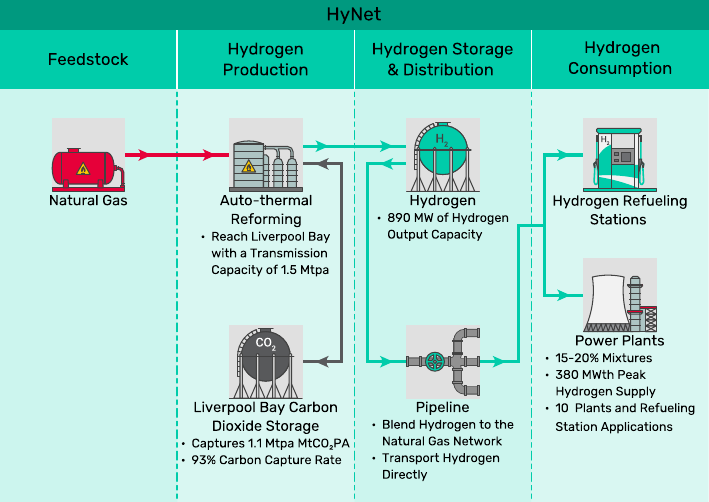}
    \caption{The overview of the HyNet project in the UK \cite{unknown2021c}.}
    \label{fig: The overview of the HyNet project in the UK}
\end{figure}

\textbf{Project overview:} The HyNet project as outlined in Figure \ref{fig: The overview of the HyNet project in the UK} is the first large-scale utilization of hydrogen in the UK \cite{HyNetproject}. This project provides a modern example of hydrogen production through Auto Thermal Reforming (ATR) of natural gas in conjunction with CCUS technology. \cite{isaac2019a}. This project aims at creating the world’s first low-cost, large-scale, and net-zero carbon emission industrial cluster \cite{hynetplant}. The total capacity of the first two hydrogen production units in the HyNet project is planned to be 890MWth \cite{robert2021a}. A portion of the hydrogen produced can be transported by pure hydrogen pipelines to 10 demonstration sites, including electric power plants and hydrogen refueling stations, and another portion can also be directed into existing local natural gas pipelines. The $\mathrm{CO_2}$
 captured can be transported to the Hamilton and Lenox natural gas fields in Liverpool Bay and stored underground \cite{hynetplant,Liverpool-Manchester,westnorthhynet}. According to reports, further benefits to the UK are £31bn of Gross Value Added (GVA) and additional employment of 11,259 job opportunities \cite{hyneteconomics}.

\textbf{Project highlights:} The HyNet project is developed and constructed using existing local energy infrastructures, supplying different local demonstration sites and blending hydrogen and natural gas to achieve lower emission levels in existing local natural gas pipelines, thus promoting the long-term reuse of natural gas networks while reducing costs. The ATR process achieves greater efficiency than Steam Methane Reforming (SMR), allowing hydrogen production to be combined with compression, thereby reducing associated costs \cite{andy2020a}. In addition, natural gas and hydrogen pipelines can reduce overall energy costs, especially in the presence of technical limitations in the electricity network, support the development of low-carbon and zero-carbon power generation infrastructure, and provide alternative options for power generation projects during periods of low power demand \cite{zhao2019a}. In the economic aspect, HyNet, with low-carbon hydrogen production, can offer large-scale, deliverable, and cost-effective decarbonization pathways for different applications in domestic, commercial, and industrial sectors \cite{badejo2024a}.

\textbf{Project analysis:} The project has two key technologies: ATR and hydrogen transportation by natural gas pipeline. ATR produces hydrogen by autothermal reforming of methane, a complex reaction combining oxidation and reduction, composed of strong exothermic and endothermic processes. Synergistic interactions among precious, non-precious, and dual-precious metals can effectively expand the catalyst's specific surface area, enhancing both its catalytic activity and stability \cite{faheem2021a}. Given that the volumetric energy density of hydrogen is roughly one-third that of methane \cite{joan2002a}, it is essential to ensure that the pressure drop through existing pipes and networks remains within acceptable limits to deliver sufficient hydrogen energy. According to the techno-economic analysis undertaken for the HyNet project, minor hydrogen network reinforcements are required to provide 100\% hydrogen energy \cite{Liverpool-Manchester}. In general, the use of natural gas pipelines for hydrogen transport can rapidly increase its distribution scale and efficiency, supporting a wider adoption of hydrogen energy \cite{wang2022a}.

\subsubsection{H2Teesside: UK’s low-carbon hydrogen production projects - Case 2}\label{subsubsec: H2Teesside: UK - based Blue Hydrogen Projects}

\begin{figure}[!ht]
  \centering
  \includegraphics[width=0.75\textwidth]{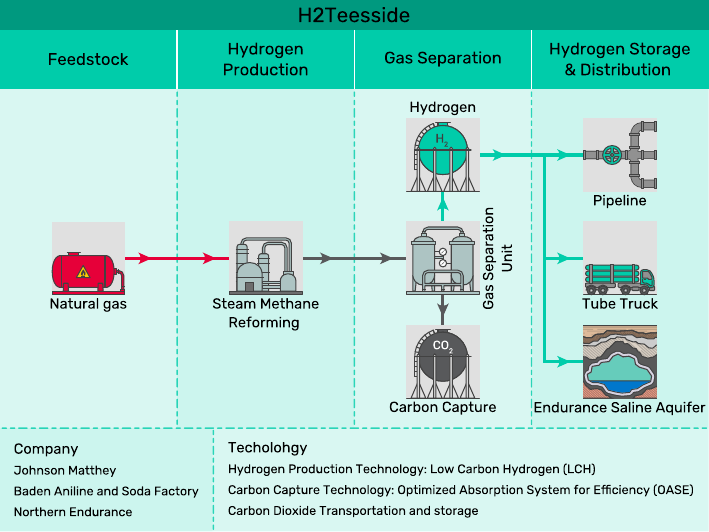}
  \caption{The overview of the H2Teesside project \cite{hydrogen-a}.}
  \label{fig: The overview of the H2Teesside project}
\end{figure}

\textbf{Project overview:} BP is planning to build the largest blue hydrogen production facility in the UK, named H2Teesside, located in Teesside, in the northeast of England. The project aims to achieve a 1.2GW hydrogen production capacity by 2030 \cite{ogwumike2024a}. The initiative aims to generate low-carbon hydrogen by transforming natural gas into hydrogen and CO\(_2\), with the CO\(_2\) being captured using CCUS. The overview of the H2Teesside project is illustrated in Figure \ref{fig: The overview of the H2Teesside project}. It is estimated that the project will capture and store up to 2 million tonnes of CO\(_2\) annually, equivalent to the heating emissions of one million UK households. H2Teesside represents a significant milestone for BP’s hydrogen business and is a critical project in supporting the UK government’s target of achieving 10GW of hydrogen production capacity by 2030 \cite{herman2024a}.

\textbf{Project highlights:} The H2Teesside project leverages the advantageous location of the Teesside region, which is close to the North Sea carbon storage sites, pipeline corridors, and existing hydrogen storage and distribution facilities. This setup not only facilitates the decarbonization of industrial clusters but also creates numerous local employment opportunities \cite{soyoung2024a}. The project aims to achieve an initial capacity of 500MW of blue hydrogen by 2027. As industrial cluster emissions decrease and hydrogen demands grow, capacity is expected to increase further by 2030. H2Teesside will also integrate with the Net-Zero Teesside (NZT) project and the Northern Endurance Partnership (NEP), both led by BP, forming one of the UK’s first carbon-neutral industrial clusters \cite{unknowp-l}.

\textbf{Project analysis:} H2Teesside provides a reliable low-carbon energy solution for industries that are difficult to electrify, serving as a crucial pillar for the UK’s net-zero emissions target \cite{ogwumike2024a}. The hydrogen produced will supply clean fuel to industrial users and households and serve as fuel for heavy transport or as a feedstock for producing biofuels and synthetic fuels. BP has also signed Memorandums of Understanding (MoUs) with several partners, including Venator, to supply clean hydrogen to its Teesside plant, and Northern Gas Networks (NGN) to drive decarbonization of the natural gas network \cite{unknowp-l}. This project will help Teesside become the UK’s first hydrogen transport hub, contributing significantly to regional economic development and global climate goals.

\subsubsection{H2FUTURE: Austria’s flagship project for green hydrogen production and metallurgical decarbonization }\label{subsubsec: H2FUTURE: Austria’s flagship project for green hydrogen production and metallurgical decarbonization}

\begin{figure}[!ht]
  \centering
  \includegraphics[width=0.75\textwidth]{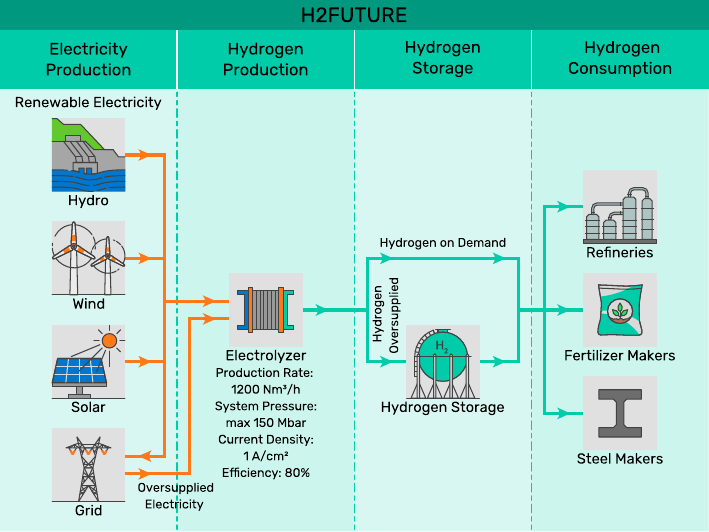}
  \caption{The overview of the H2FUTURE project in Austria \cite{h-a,unknown-g}.}
  \label{fig: The overview of the H2FUTURE project in Austria}
\end{figure}

\textbf{Project overview:} In the future, green hydrogen will lay the foundation for developing and decarbonizing the steel, refinery, fertilizer, and other industrial sectors that will demand large volumes of hydrogen \cite{pathak2023a}. The H2FUTURE project aimed to significantly reduce carbon emissions in steel production, helping Austria work toward its 2040 carbon neutrality climate target \cite{unknown2021d}. As shown in Figure \ref{fig: The overview of the H2FUTURE project in Austria}, the H2FUTURE project utilized low-carbon and zero-carbon energy resources to produce green hydrogen by water electrolysis. Furthermore, this project exploited a PEM electrolyzer as a modern, efficient, and effective technology with a capacity of 6MW, producing hydrogen at a rate of 1200m³/h. 

\textbf{Project highlights:} The H2FUTURE project was a successful case of clean energy innovation within the EU. The project highlighted two aspects: the use of PEM electrolysis technology and hydrogen metallurgy technology. In the H2FUTURE project, the PEM electrolysis technology, designed and provided by Siemens, was employed with a high efficiency exceeding 80\%. Due to the utilization of highly efficient and available equipment in the H2FUTURE project, the cost of producing green hydrogen was remarkably low \cite{unknown-g,h-b}. Additionally, the project adopted two hydrogen metallurgy technology routes: (i) hydrogen-rich reduction technology based on an existing gas-based reduction process and (ii) blast furnace injection hydrogen-rich technology based on blast furnace ironmaking. This almost covered the current processes of ironmaking and steel production \cite{unknown-h}. The hydrogen metallurgy technology of H2FUTURE can significantly reduce carbon emissions and support decarbonization in steel production, making it promising.

\textbf{Project analysis:} The key technology used in the project was a PEM electrolyzer. The project employed the Siemens SILYZER electrolyzer. From a technological standpoint, PEM electrolyzers can mitigate potential issues related to the contamination and corrosion of ALK electrolyzers, offering a higher lifespan. Furthermore, PEM electrolyzers are characterized by a lightweight, compact structure, 99.99\% hydrogen purity, and fast responses \cite{carmo2013a}. Leveraging this technology, the electrolysis process in PEM electrolyzers can be seamlessly scaled up and promptly adapted to operational conditions, making it highly suitable for dynamic industrial processes and grid services. In addition, it had been implemented to automatically provide a Frequency Control Reserve (FCR) by adjusting the power consumption of the electrolyzer based on local grid frequency measurements \cite{unknown-i}.

In hydrogen metallurgy, hydrogen serves as a reducing agent during the iron ore reduction smelting process \cite{zhang2022a}. Furthermore, hydrogen metallurgy can be categorized into hydrogen-rich metallurgy (mixed gas containing CO and hydrogen) and pure hydrogen metallurgy \cite{tang2020a}. Hydrogen metallurgy technology substitutes hydrogen for coal, and when hydrogen reduces iron oxide, the process generates metallic iron and water vapor, significantly lowering $\mathrm{CO_2}$ emissions in the steel sector. Moreover, the use of hydrogen noticeably enhances the reduction efficiency of iron ore, thereby boosting overall efficiency. As an exemplary project embracing hydrogen metallurgy technology, H2FUTURE has provided a background for the development of hydrogen metallurgy, offering insights for other industrial sectors that require hydrogen.

\subsubsection{Demo4GRID: European project for green hydrogen production and grid interaction}\label{subsubsec: Demo4GRID: European project for green hydrogen production and grid interaction}

\begin{figure}[!ht]
  \centering
  \includegraphics[width=0.75\textwidth]{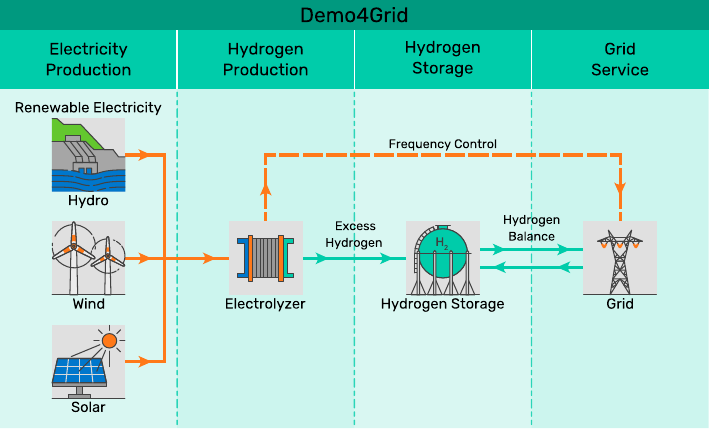}
  \caption{{The features of the Demo4Grid project} \cite{unknown-j,unknown-k}.}
  \label{fig: The features of the Demo4Grid project}
\end{figure}

\textbf{Project overview:} The Demo4Grid project was located in Innsbruck, Austria. It was jointly funded by the Swiss government and the EU, with a total budget of €7.7m \cite{unknown-k}. The partners of the Demo4Grid project included Diadikasia Symvouloi Epicheiriseon in Greece, IHT in Switzerland, Aragon Hydrogen Foundation and Instrumentation Componentes SA in Spain, and MPREIS Warenvertriebs GmbH and FEN Sustain Systems GmbH in Austria \cite{unknown-k}. This project enables the storage of excess electricity from renewable energy, such as solar, wind, and hydropower, by converting it into hydrogen via electrolyzers, as shown in  Figure \ref{fig: The features of the Demo4Grid project}. The electrolyzer could provide grid frequency control services to ensure power balance, and hydrogen could be converted back into electricity again when the grid needed it. This function can boost renewable energy usage and lower carbon emissions \cite{unknown-k,unknown-l}.

\textbf{Project highlights:} The Demo4Grid project implemented the EU’s largest single-stack ALK electrolyzer, providing grid frequency control services. It utilized the electrolyzer not only to generate green hydrogen but also to ensure grid frequency stability when renewable energy was connected to the grid. In this project, surplus electricity from intermittent renewable sources, such as wind or solar, can be stored within the system as hydrogen. At low grid frequencies, the system can reconvert to electricity, enhancing renewable energy utilization and reducing carbon emissions.

\textbf{Project analysis:} The ultimate goal of the Demo4Grid project was to offer grid-balancing ancillary services (primary and secondary frequency regulation services) to the transmission system operator, Austrian Power Grid (APG), to generate more economical hydrogen \cite{unknown2017b}. The project combined the ALK electrolyzer with hydrogen storage tanks to stabilize the frequency of the grid. In practice, both ALK and PEM electrolyzers provide a sub-second response and, thereby, are capable of offering ancillary services to reduce costs and durations of frequency disturbances in the grid. Both electrolysis technologies can respond quickly to stabilize the grid when the electrolysis system is operating at operating temperature \cite{unknown-m,unknown-n}.

After a power generation outage, the power system frequency will change. The maximum instantaneous frequency deviation where the frequency occurs immediately is allowed to be 0.8Hz below the nominal frequency \cite{duo-a}. The maximum frequency deviation is relative to the boundary where stability is maintained. Load Frequency Control (LFC) is used to restore the frequency to this range by adjusting the power reserve. These reserves can be adjusted, increasing when power is capped and decreasing when resources are sufficient \cite{duo-a}. In Europe, Frequency Regulation Reserve (FCR) and Frequency Restoration Reserve (FRR) are widely used, and controllable loads on the demand side are also eligible to participate in these grid services through dynamic power regulation \cite{duo-a}.

In response to a disturbance, first, FCRs are activated to stabilize the frequency deviation at a point away from the nominal level. Then, FRRs engage to restore the frequency to its nominal level. The time taken to adjust reserves influences the duration required for frequency stabilization. Presently, the Full Activation Time (FAT) for FCR, or the full reserve deployment time, is 30 seconds, and FRR must activate within a 15-minute window \cite{duo-a}. 

\subsubsection{H2 Green Steel: Swedish hydrogen-based green steel project}\label{subsubsec: H2 Green Steel: Swedish Hydrogen - based Green Steel Project}

\begin{figure}[!ht]
  \centering
  \includegraphics[width=0.75\textwidth]{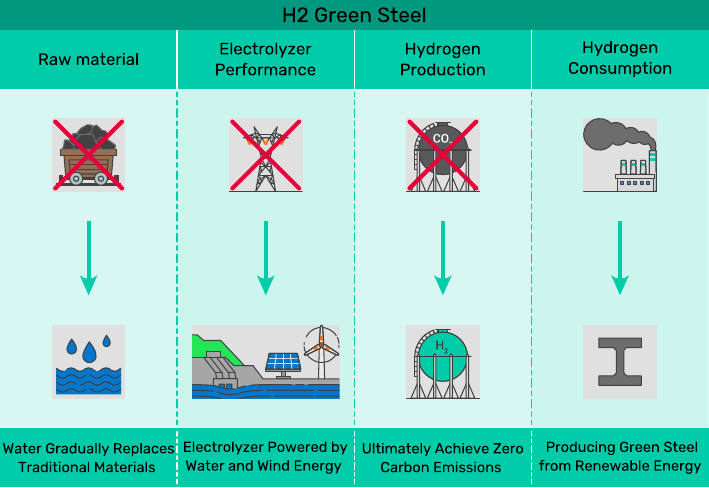}
  \caption{The overview of the H2 Green Steel.}
  \label{fig: The overview of the H2 Green Steel}
\end{figure}

\textbf{Project overview:} H2 Green Steel is a large-scale green hydrogen project based in Sweden that aims to produce green steel using renewable energy \cite{Elmfeldt2024}. The overview of the H2 Green Steel project is illustrated in Figure \ref{fig: The overview of the H2 Green Steel}. The project involves electrolyzing water to hydrogen and oxygen, utilizing hydrogen to replace traditional coal and natural gas as a reducing agent in steel production, significantly reducing $\mathrm{CO_2}$ emissions \cite{unknowp-n}. The ultimate goal is zero-emission steel production, driving the global steel industry toward a greener and more sustainable future. It aims to become one of Europe’s largest hydrogen-based steel production initiatives.

\textbf{Project highlights:} H2 Green Steel plans to construct a 700MW electrolyzer, making it one of the most significant green steel projects worldwide. The electrolyzer will directly supply hydrogen for steel production. The project has received support from Vattenfall and Iberdrola, with funding from the Swedish government and several European financial institutions \cite{unknowp-n}. The project will leverage renewable energy sources such as wind and hydropower to ensure zero-carbon emissions during production \cite{unknowp-o}. It is projected to reduce approximately 5 million tons of $\mathrm{CO_2}$ emissions annually, profoundly impacting the economy of Sweden and global environmental efforts \cite{Ledari2023}.

\textbf{Project analysis:} The core innovation of H2 Green Steel lies in using hydrogen to replace coal as a reducing agent, significantly cutting $\mathrm{CO_2}$ emissions. This technology is highly innovative globally \cite{Elmfeldt2024,unknowp-o}. As the demand for low-carbon solutions increases, the market for green steel is expanding. However, the project faces challenges such as high costs for building and operating electrolyzers, substantial capital investment, and the sustainability of the hydrogen supply chain. Its success will improve Sweden's competitiveness in the steel industry and drive global decarbonization efforts. Its sustainability depends on the stability of the green electricity supply, market acceptance, and policy support.

\subsection{North American demonstration projects for hydrogen production}\label{subsec: North American demonstration projects for hydrogen production}

\subsubsection{Utah hydrogen power project}\label{subsubsec: Utah Hydrogen Power Project}

\begin{figure}[!ht]
  \centering
  \includegraphics[width=0.75\textwidth]{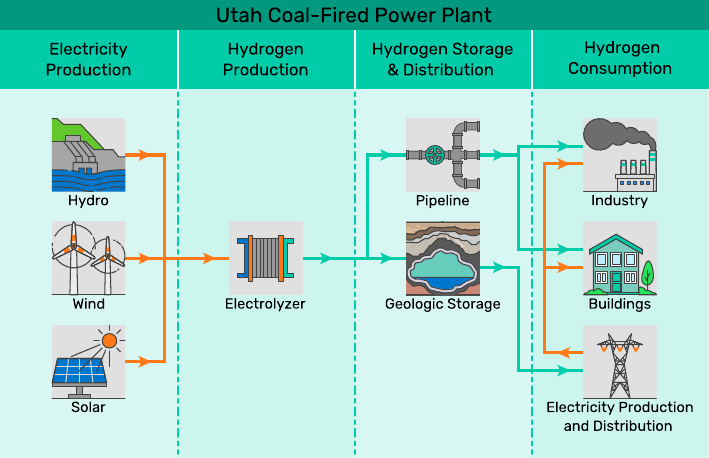}
  \caption{Utah coal-fired power plant in the United States \cite{storage-a}.}
  \label{fig: Utah coal-fired power plant in the United States}
\end{figure}

\textbf{Project overview:} The Intermountain Power Project (IPP) is a 1.8GW coal-fired power plant located in Utah, United States \cite{unknown-p,unknown-o}. Since 1986, it has been serving the Southern California Public Power Bureau, as well as small municipal and cooperative utility units in Nevada and Utah \cite{unknown-p}. In recent years, it has been transitioning toward renewable energy generation. Without hydrogen, it is impossible to achieve 100\% renewable energy. On March 10, 2020, Mitsubishi Hitachi Power Systems announced a contract with Intermountain Power Agency (IPA) to provide advanced gas turbines \cite{unknown-o} aimed at promoting the transition of a power plant in central Utah from fossil fuel to renewable fuel, as shown in Figure \ref{fig: Utah coal-fired power plant in the United States}. By 2045, the ultimate goal is to rely entirely on hydrogen for carbon-free power generation \cite{unknown-o}. The updated facility will provide 840MW of electricity to Los Angeles and other cities served by the IPA in California and Utah in the future \cite{storage-a}.

\textbf{Project highlights:} Salt cavern hydrogen storage is a major feature of this project, with Utah having the largest underground salt layer and outstanding underground gas storage capacity \cite{unknown-q}. More than 100 caves can be built near this project, each larger than the Empire State Building. Salt caverns can store renewable hydrogen and compressed air, enabling the storage of electricity produced from renewable energy. Salt caverns are ideal for large-scale hydrogen storage, with relatively low costs, while other large-scale hydrogen storage technologies have higher costs \cite{miocic2023a}. 

Therefore, hydrogen storage is one of the unique characteristics of this project. This approach facilitates seasonal storage of renewable energy by converting electricity into hydrogen and storing it underground for long periods \cite{miocic2023a}.

\textbf{Project analysis:} This project serves as a significant demonstration of transitioning traditional coal-fired power plants to zero-emission energy systems by incorporating green hydrogen power generation \cite{renewed-a}. Utilizing alkaline electrolysis driven by solar and wind energy, the project can provide stable, dispatchable, clean energy to Southern California while reducing dependence on natural gas storage. Through low-cost hydrogen storage, hydrogen power generation, and energy diversification, the project not only significantly reduces greenhouse gas emissions but also establishes a new benchmark for regional economic development and energy system sustainability.

{\subsubsection{H2Hubs: The US clean hydrogen hubs initiative}\label{subsec: H2Hubs: The US Clean Hydrogen Hubs Investment Initiative}


\begin{figure}[!ht]
  \centering

  \begin{subfigure}[b]{0.75\textwidth}
    \centering
    \includegraphics[width=\textwidth]{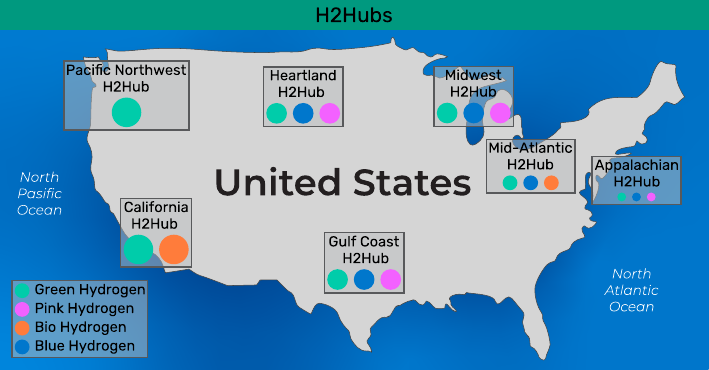}
    
    \label{fig:h2hubs-a}
  \end{subfigure}

  \begin{subfigure}[b]{0.75\textwidth}
    \centering
    \includegraphics[width=\textwidth]{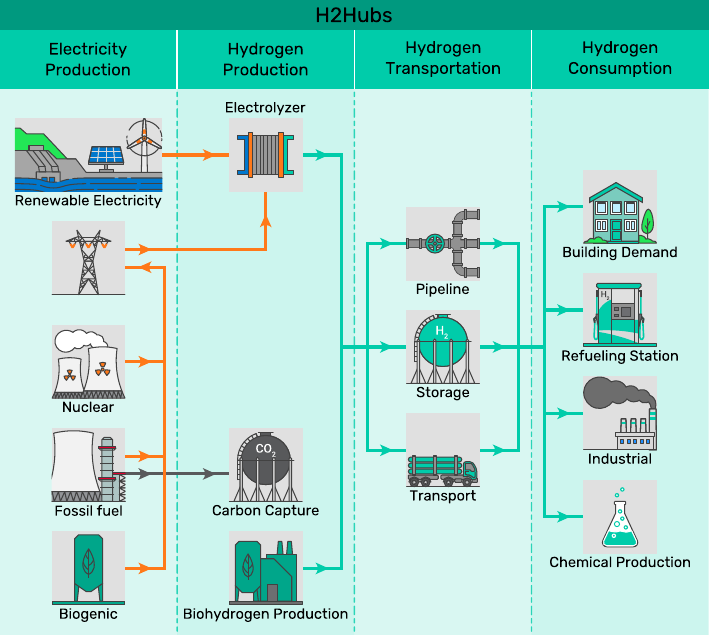}
   
    \label{fig:h2hubs-b}
  \end{subfigure}

  \caption{The overview of the H2Hubs project \cite{bade2024a,unknowp-m}.}
  \label{fig: The overview of the H2Hubs project}
\end{figure}

\textbf{Project overview:} The US Department of Energy (DOE) is investing \$7 billion to establish seven Regional Clean Hydrogen Hubs (H2Hubs) across the country. As shown in Figure \ref{fig: The overview of the H2Hubs project}, these hydrogen hubs will connect clean hydrogen producers, consumers, and related infrastructure to support a full supply chain from production to end-use \cite{bade2024a}. The H2Hubs span 16 states and are projected to produce 3 million metric tons of hydrogen annually, accounting for one-third of the 2030 clean hydrogen production target \cite{unknowp-m}.  The initiative aims to accelerate decarbonization in different energy sectors, supporting the US goal of achieving net-zero carbon emissions by 2050 while fostering regional economic growth.

\textbf{Project highlights:} The highlights of the H2Hubs project lie in its regional adaptability and technological diversity. The hydrogen hub along the Gulf Coast leverages abundant natural gas resources and CCUS facilities to produce blue hydrogen while integrating renewable energy to produce green hydrogen. The Midwest hydrogen hub relies on nuclear power to provide a stable energy source for large-scale green hydrogen production. Meanwhile, the California hydrogen hub primarily uses water electrolysis to produce green hydrogen, meeting local clean energy demands \cite{unknowp-m}. In addition to reducing greenhouse gas emissions, the H2Hubs project aims to create tens of thousands of high-quality jobs, boosting economic recovery and regional development.

\textbf{Project analysis:} The H2Hubs project exemplifies the close integration of clean energy technology and economic policy \cite{bade2024a}. By consolidating regional resources and implementing phased investments, H2Hubs significantly reduces the risks of deploying large-scale hydrogen technology. The project emphasizes technological innovation and optimization of existing infrastructure, such as hydrogen pipelines and carbon storage facilities \cite{unknowp-m}. Furthermore, H2Hubs accelerates the development of a large-scale clean hydrogen market by integrating with the industrial, transportation, and power generation sectors, providing a low-carbon flexible solution for future energy systems \cite{brunik2024a}. Implementing H2Hubs will reinforce the US's leadership in the global hydrogen market, establishing it as a flagship project in the energy transition.}

{\subsubsection{EverWind Fuels: Canadian green hydrogen industry promotion project}\label{subsec: EverWind Fuels: Canadian Green Hydrogen Industry Promotion Project}

\begin{figure}[!ht]
  \centering
  \includegraphics[width=0.75\textwidth]{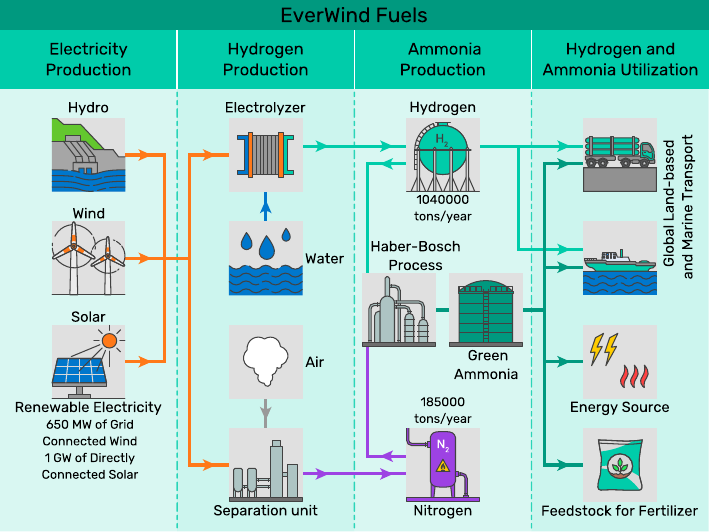}
  \caption{The overview of the EverWind Fuels project \cite{VanPraet2023}.}
  \label{fig: The overview of the EverWind Fuels project}
\end{figure}

\textbf{Project overview:} EverWind Fuels, located in Point Tupper, Nova Scotia, Canada, is North America's first industrial-scale green hydrogen and green ammonia production facility \cite{Graney2023a}. As shown in Figure \ref{fig: The overview of the EverWind Fuels project}, the project primarily involves wind and solar power generation supplying energy to PEM electrolyzers for green hydrogen production and green ammonia synthesis. The project is being developed in two phases with a total investment of approximately \$13.7 billion. The first phase involves an investment of about \$3.4 billion, with installed capacities of 650MW of grid-connected onshore wind and 300MW of directly connected solar PV, supporting a 280MW PEM electrolyzer. It is expected to produce approximately 40,000 metric tons of green hydrogen and 240,000 metric tons of green ammonia annually, with operations planned to commence in early 2026. The project received environmental approval from the Nova Scotia provincial government in February 2023 and plans to export products to international markets, including Germany. The project is expected to make significant contributions to Canada’s clean technology industry and local economy, create over 820 full-time equivalent jobs, and provide more than \$30 million annually in grid services. \cite{VanPraet2023}.

\textbf{Project highlights:} EverWind Fuels utilizes renewable energy-driven electrolysis technology to produce zero-emission green hydrogen and green ammonia. Its strategic location near the Port of Point Tupper ensures efficient international transportation of these products. The project complies with Canadian and international energy and environmental policies, gaining robust support from local governments and environmental organizations \cite{VanPraet2023}. By collaborating with German energy firms to penetrate European markets, it not only sets a benchmark for hydrogen energy development in North America, but also accelerates the global energy transition and contributes significantly to the advancement of the hydrogen economy.

\textbf{Project analysis:} The EverWind Fuels project demonstrates the industrial potential of green hydrogen and ammonia by integrating advanced technologies, renewable energy sources, and strategic international partnerships. It offers a practical solution for reducing global carbon emissions and meeting the growing demand for clean energy. In addition, the project supports local economic growth by creating hundreds of jobs and enhancing the regional infrastructure \cite{unknowp-p}. However, challenges such as ensuring a stable renewable energy supply, managing the costs of hydrogen storage and transportation, and addressing competition in international markets remain critical. In general, the EverWind Fuels project represents a significant step toward the commercialization of hydrogen energy and positions North America as a key player in the global energy transition.}

\subsection{Asian demonstration projects for hydrogen production}\label{subsubsec: Asian demonstration projects for hydrogen production}

\subsubsection{Asian Renewable Energy Hub (AREH)}\label{subsubsec: Asian Renewable Energy Hub (AREH)}

\begin{figure}[!ht]
  \centering
  \includegraphics[width=0.75\textwidth]{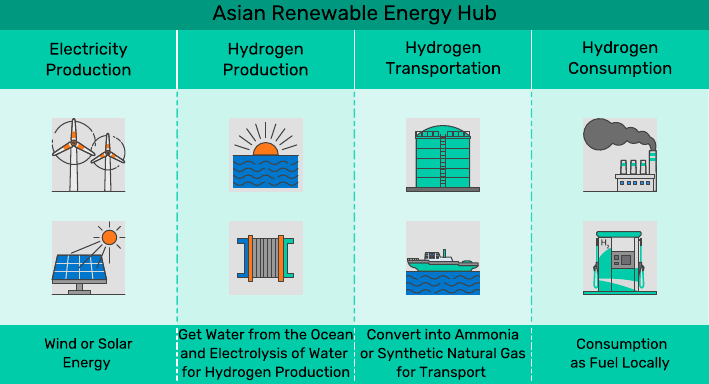}
  \caption{The overview of the AREH project \cite{unknown-w}.}
  \label{fig: The overview of the AREH project}
\end{figure}

\textbf{Project overview:} Figure \ref{fig: The overview of the AREH project} presents the schematic diagram of this project. The Asian Renewable Energy Hub (AREH), located in Western Australia’s desert region, is a green hydrogen initiative planning to construct a hybrid power plant featuring wind turbines, solar panels, and one of the largest centers for green hydrogen and green ammonia production worldwide \cite{unknown2024a,unknown-t,unknown-u}. The project will be developed on a 6,500\SI{}{\kilo\metre\squared} site in the Pilbara region near the Indian Ocean \cite{sciences-a}. After completion, the project, at total capacity,  is projected to generate 1.6 million tons of green hydrogen and 9 million tons of green ammonia yearly, with plans for export \cite{unknown-v}.

\textbf{Project highlights:} The AREH project will be powered by 26GW of wind and solar energy \cite{bp-a}, including 16GW of onshore wind and 10GW of photovoltaics, powering 14GW of electrolyzers to produce renewable hydrogen and ammonia \cite{unknowp-q}. The AREH, with a proposed installed capacity of 26GW, setting it up to be one of the largest global renewable energy hubs. This project is geared toward producing significant quantities of green hydrogen and ammonia, marking a substantial step in the global transition to clean energy.

\textbf{Project analysis:} In its initial phase, the AREH project will include a 7.5GW wind power and a 3.5GW solar power installation, creating a total capacity of 11GW \cite{unknown-x}. Construction of the first phase will be finished in 2025 \cite{unknown-x}. The wind power generation component will account for two-thirds of the project’s installed capacity. 3GW of electricity will serve local industry, and much of the remaining energy will be used to create green hydrogen and green ammonia, which will then be exported by sea \cite{unknowp-v}. The latter will be used in power generation, pharmaceuticals, ship fuel, and fertilizer \cite{yuezba2022a}. This project’s vast scale will drive the establishment of new supply chain facilities for the manufacturing and assembly of wind, solar, and hydrogen production equipment, generating significant local employment opportunities. Approximately 20,000 jobs will be created throughout the 10-year construction phase, along with 3,000 long-term roles for its 50+ years of operation \cite{unknown-y}. The annual electricity generation of the AREH project is expected to exceed 90TWh \cite{unknowp-w}, supplying both domestic and international markets. To underscore the significant scale of the project, it is notable that this figure represents approximately one-third of Australia’s total electricity generation, which stood at 272 TWh in the previous year.\cite{unknown-z}.

\subsubsection{FH2R: Japan’s Fukushima photovoltaic hydrogen production project}\label{subsubsec: FH2R: Japan’s Fukushima photovoltaic hydrogen production project}

\begin{figure}[!ht]
  \centering
  \includegraphics[width=0.75\textwidth]{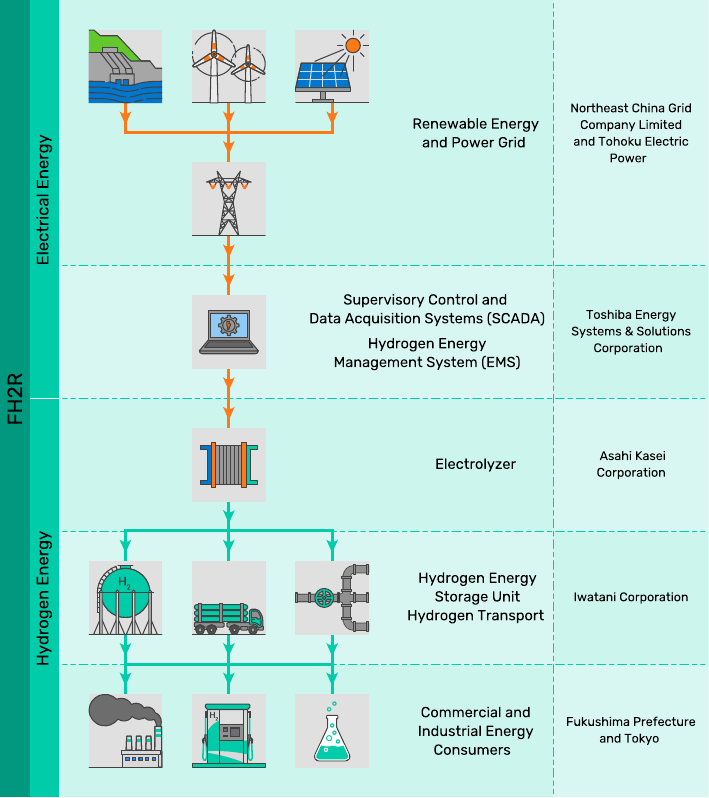}
  \caption{The overview of the FH2R project \cite{unknowo-a}.}
  \label{fig: The overview of the FH2R project}
\end{figure}

\textbf{Project overview:} The diagram of the project is depicted in Figure \ref{fig: The overview of the FH2R project}. The FH2R project is a typical representative of hydrogen source autonomy in Japan’s hydrogen energy strategy \cite{khan2022a}. The FH2R project focuses on large-scale hydrogen production, storage, and distribution, stabilizing power grid supply and demand and promoting its practical application. In the FH2R project, Toshiba spearheaded the development of a hydrogen energy management system and oversaw the entire project’s construction. Meanwhile, Tohoku Electric Power concentrated on monitoring the energy management system, implementing Supervisory Control and Data Acquisition (SCADA) systems, and adhering to grid regulations. Iwatani took on the task of investigating hydrogen supply and demand forecasting systems alongside managing hydrogen transportation and storage \cite{unknowo-b,unknowo-c}. The ALK electrolyzer provided by Asahi Kasei Corporation had advanced chlor-alkali electrolysis technology \cite{unknowo-d}. On the one hand, the FH2R project managed hydrogen production and storage in alignment with a forecasting system tailored to downstream market demand and supply. Furthermore, it also addresses grid adjustment requirements by regulating the hydrogen production unit.

The distances of the FH2R project from Fukushima Prefecture, Sendai City, and Tokyo City are 70km, 100km, and 250km, respectively. Equipped with 20MW photovoltaic power generation, the maximum power of the electrolyzer is 10MW, the rated power is 6MW, and the hydrogen production capacity (rated working condition) is 1200 Nm³/h \cite{unknowo-e}. The project spans 220,000\SI{}{\metre\squared}, with 180,000\SI{}{\metre\squared} dedicated to the photovoltaic field and 40,000m allocated for research and facilities. 

\textbf{Project highlights:} The FH2R project uses P2G technology. As production scaled up, large amounts of hydrogen energy could be stored, building a new supply chain to optimize hydrogen’s potential. The project also uses a hydrogen energy management system to optimally utilize hydrogen production and storage facilities to ensure the dynamic balance of grid supply and demand and to implement and investigate effective and efficient operational control technologies. The technology combined demand response from the grid with supply and demand response from hydrogen using equipment with different operating cycles. In terms of control, the system achieved overall control and dynamic balance through the operation schedule function and control function of the electrolyzer and power-conditioning system for solar power.

\begin{figure}[!ht]
  \centering
  \includegraphics[width=0.75\textwidth]{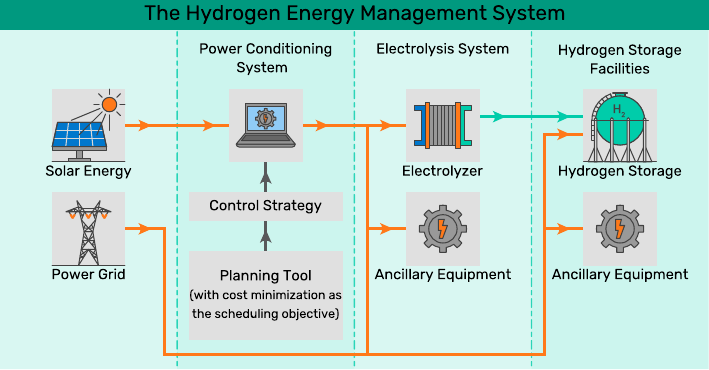}
  \caption{The diagram of the hydrogen energy management system \cite{unknowo-c}.}
  \label{fig: The diagram of the hydrogen energy management system}
\end{figure}

\textbf{Project analysis}: The FH2R project aims to optimize demand response through the grid, maximize grid utilization without the use of storage batteries, and establish low-cost green hydrogen production technologies. Furthermore, electrolyzers resolve the intermittency issue of solar power, allowing for adjustments aligned with grid supply and demand. The hydrogen produced by the project powers stationary hydrogen fuel cell systems, fuel cell vehicles, and buses. The EMS is outlined in Figure \ref{fig: The diagram of the hydrogen energy management system}, utilizing the operation schedule function to plan the operation schedule for each unit and notifying the control function of each unit. The control function mainly includes the electrolyzer management and power-conditioning system for solar power. Through these two functions, the system sends optimal control actions to various devices and units.

\subsubsection{Ordos zero-carbon smart green hydrogen production project}\label{subsubsec: Ordos smart zero-carbon green hydrogen production project}

 \textbf{Project overview:} The Ordos zero-carbon smart green hydrogen production project relies on photovoltaic and storage facilities and the advanced water electrolysis hydrogen production system, to carry out the first photovoltaic off-grid hydrogen production demonstration in China \cite{unknowo-n}. This project is set to foster integrated growth between the conventional synthetic materials chemical industry and the hydrogen energy sector \cite{unknowo-f}. The project uses the advanced water electrolysis hydrogen production system, “Hydrogen Boat,” independently developed by Haide Hydrogen Energy, which can be directly connected to the on-site solar power station to directly produce green hydrogen \cite{group2023a,unknown2023a}.

\textbf{Project highlights:} The Hydrogen Boat, a standardized pressurized PEM water electrolysis hydrogen production system independently developed by Haide Hydrogen Energy, has undergone technological innovations from critical components and mechanical design to electrolysis stacks and control systems. The system integrates three major green hydrogen technologies: Haide SENSOR 2.0, Dongao’s hydrogen safety technology, Yunfan electrolytic hydrogen production monolithic, and the hydrogen production digital platform COMPASS, to effectively solve the issues currently plaguing the green hydrogen manufacturing industry and provide safety efficiency, and intelligent advanced green hydrogen production solutions \cite{unknowo-g}.

\textbf{Project analysis:} The Hydrogen Boat system achieves substantially higher efficiency and adaptability to green power than the average system in this industry. Electricity, as the main expense, makes up about 80\% of the total cost in these systems \cite{unknowp-s}. The DC power consumption of the Hydrogen Boat is 4.0-4.3kWh/Nm³, which is far lower than the industry average indicator of 4.3-4.8kWh/Nm³ \cite{unknowp-t}. It achieves a measured power consumption of 3.98kWh/Nm³, and the hydrogen production efficiency reaches more than 80\% \cite{unknowp-t}. On the start-up side of hydrogen production, the Hydrogen Boat uses self-developed new energy system control technology to increase the start-up speed and reduce losses when connecting to green electricity. On the hydrogen production operation side, based on the local fluctuation of solar power, a coupling control strategy ensures safe operation, lowering the limit of hydrogen production system load and reducing energy consumption \cite{unknowo-j}.

The Hydrogen Boat introduces a dynamically expandable new generation of standardized square hydrogen production units called ``Yunfan." Yunfan’s thin square structure can improve the uniformity and consistency of multi-physics fields at the monolithic level compared to traditional circular electrolytic monoliths \cite{unknowo-k}. It has a more uniform flow field division with less impact on the diaphragm, allowing the electrode to react more thoroughly, thus increasing the current density. A further innovation is that Yunfan can also be expanded. Using Yunfan’s scalability, Hydrogen Boat can expand output from 50Nm³/h to 500Nm³/h. For the solution within the hydrogen production plant, Yunfan can expand the capacity of a single tank to 1,000Nm³/h \cite{unknowp-u}.

In addition, the plug-in structure with piezoelectric stack design enables in-situ replacement of a single chip within 12 hours, significantly reducing operation and maintenance costs \cite{unknowp-t}. This stack technology structure boosts hydrogen production efficiency and reduces ongoing operation and maintenance costs effectively.

About dynamic performance, this system exhibits stability over a wider load range, operating seamlessly from 20\% to 110\% of its designed capacity \cite{unknowo-l}. The system loading/unloading rate reaches 5\% per second, which is much better than the industry average. For green hydrogen production conditions, advanced intelligent sensing and control technology is used to achieve temperature stability and pressure stability under fast load conditions, ensuring the hydrogen production system’s safety while enhancing the overall efficiency of hydrogen production \cite{unknowo-i}. Haide wireless hydrogen safety sensor can monitor hydrogen leakage in real-time and minimize the risk of explosion. Yunfan’s single chip integrates a tracking system. Through intelligent Cloud Virtual Machine (CVM) detection technology, real-time inspection and collection of information on every single chip, real-time tracking of equipment operation, and creation of a hydrogen production digital cloud platform, this hydrogen production system advances digitalization and management of hydrogen systems. Once a fault occurs, maintenance only requires the replacement of the fault reporting unit without the need to transport the entire machine back to the site for repair. The maintenance time of the whole equipment is shortened from weeks to hours. In the photovoltaic hydrogen production scenario, the overall economic performance is improved by 25\% \cite{unknowp-s}.

\subsubsection{Ningxia Ningdong project for renewable hydrogen production and carbon emission reduction}\label{subsubsec: Ningxia Ningdong project for renewable hydrogen production and carbon emission reduction}

\begin{figure}[!ht]
  \centering
  \includegraphics[width=0.75\textwidth]{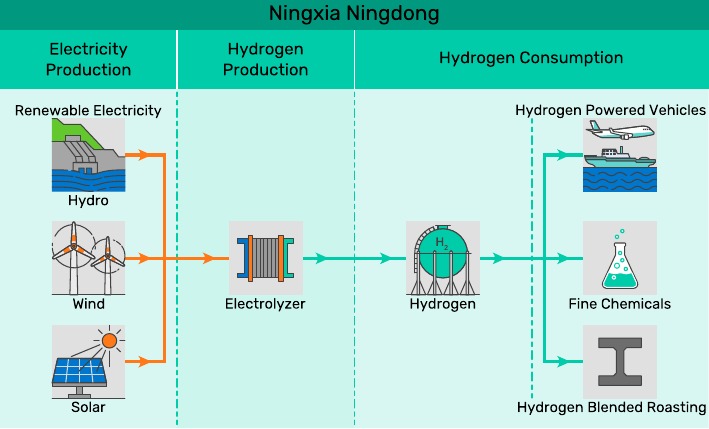}
  \caption{The overview of the Ningxia Ningdong project for renewable hydrogen production and carbon emission reduction \cite{unknowo-s}.}
  \label{fig: NINGXIA}
\end{figure}

\textbf{Project overview:} As shown in Figure \ref{fig: NINGXIA}, the Ningxia Ningdong project for renewable hydrogen production and carbon emission reduction is China’s first commercialized renewable hydrogen project integrated with the chemical industry and transportation scenarios. It is Ningxia’s first hydrogen energy complete industry chain innovative ecological project. The project integrates production, storage, processing, and utilization with a total investment of ¥3.8 billion \cite{unknowo-q,unknowo-r}. Key construction projects include a 620MW photovoltaic system, two hydrogen production facilities with a capacity of 20,000Nm³/h, two hydrogen refueling stations with a total capacity of 4,000kg/d, and associated hydrogen pipelines. It is expected to produce 4,509 tons of high-purity hydrogen per year, and the hydrogen produced will be used by hydrogen-powered trucks transporting coal and coal-to-methanol production in coal chemical industry parks \cite{unknowp-r}.

\textbf{Project highlights:} Part of the hydrogen produced will supply the coal-to-liquids project of Ningxia Coal Industry, realizing partial green hydrogen replacement, and part will supply hydrogenation stations to complete the hydrogen refueling work of hydrogen heavy trucks transporting coal between surrounding thermal power plants and coal mines. After completion and operation, the average annual power generation of the photovoltaic station will be 983GW, saving an average of 334,400 tons of standard coal and reducing emissions by 901,200 tons per year. Furthermore, it is highly significant for promoting energy transition, maximizing clean energy use, and supporting low-carbon economic growth at Ningdong Energy and Chemical Base, China’s largest coal-to-liquid chemical center \cite{unknowo-t}.

\textbf{Project analysis:} The facilities consist of a 15,000Nm³/h Qingshuiying hydrogen production station, a 5,000Nm³/h Wynn hydrogen production station, a 1,000kg/day Zaoquan hydrogenation station, and a 5,000kg/day Yongli hydrogen refueling station \cite{unknowo-u}. Qingshuiying Hydrogen Production Station is located in Ningdong, with a total investment of ¥377.6 million, an environmental protection investment of ¥3.01 million, a hydrogen production scale of 15,000Nm³/h, and a designed annual operating time of 2,300 hours \cite{shi2023a}. A total of 16 alkaline electrolyzers will be installed \cite{unknowo-v}. The Wynn hydrogen production station is equipped with supporting construction of a circulating water system, hydrogen storage tank, accident oil tank, accident pool, septic tank, and other public auxiliary and environmental protection projects, with a hydrogen production scale of 5,000 standard cm/hour. Benefiting from 3,000 hours of sunshine annually, Ningdong has ample solar and hydrogen resources, with green hydrogen production costing less than \$3 per kg \cite{unknowo-w}. The Zaoquan hydrogenation station is equipped with two hydrogen storage bottle groups with a total hydrogen storage capacity of 778.68kg, two hydrogen compressors, two hydrogenation machines, one nitrogen bottle group (16 pieces), and two hydrogen long tube trailers.

This project has developed an extensive industrial chain that includes renewable hydrogen production, the substitution of coal-based chemicals with green hydrogen, hydrogen storage and transportation, and hydrogen refueling services \cite{unknowo-x}. The Yongli hydrogen refueling station includes six 45MPa tube-bundled hydrogen storage units, with a combined capacity of 1,557.36kg. In the future, two additional 87.5MPa tube-bundled storage units are reserved, providing an extra capacity of 367.84kg \cite{unknowo-t}. There are two hydrogen long-tube trailers. Each hydrogen long-tube trailer holds 24m³ of water and can store up to 700kg of hydrogen. The electric energy used in the hydrogenation station comes from photovoltaic green electricity, and the hydrogen produced is green and environmentally friendly.

Through the construction of large-scale green hydrogen bases, innovation in key hydrogen energy technologies will be promoted, and the construction of a holistic industrial ecosystem will be accelerated, effectively promoting the adjustment of Ningdong’s coal-heavy energy structure and high-carbon industrial structure and accelerating the transition toward modern energy systems.

\subsubsection{Chifeng project for wind-solar-based green hydrogen and ammonia production}\label{subsubsec: Chifeng project for wind-solar-based green hydrogen and ammonia production}

\begin{figure}[!ht]
  \centering
  \includegraphics[width=0.75\textwidth]{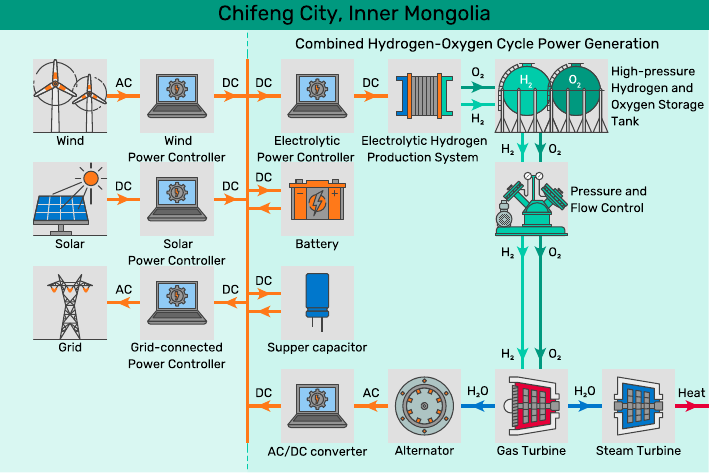}
  \caption{{The green hydrogen and ammonia production from wind and solar energy in Chifeng City, Inner Mongolia \cite{unknowp-h}.}}
  \label{fig: CHIFENG}
\end{figure}

\textbf{Project overview:} The green hydrogen and ammonia demonstration project in Chifeng, illustrated in Figure \ref{fig: CHIFENG}, represents a collaborative effort between Envision Energy and Chifeng State-owned Assets Group, with a total investment of around ¥40 billion \cite{unknowp-f}. This project aims to produce 1.52 million tons of carbon-free industrial gas products yearly. It will be developed in three phases and is scheduled for full completion and operational status by 2028 \cite{unknowp-g}. Upon completion, the facility will supply zero-carbon ammonia, hydrogen, oxygen, nitrogen, and other industrial gases, contributing substantially to carbon reduction. Leveraging the strengths of zero-carbon gas products and a state-of-the-art energy storage air separation industry, Yuanbaoshan Chemical Industry Park is positioned to evolve into a zero-carbon chemical industry hub.

\textbf{Project highlights:} This project, developed by Envision Technology in partnership with the Government, was formalized through a strategic agreement signed in March 2022. By leveraging technological innovation, wind power and energy storage are envisioned as the “new coal,” while the zero-carbon industrial park represents the “new infrastructure.” The project aims for an annual production of 1.52 million tons of carbon-free industrial gases \cite{unknowp-i}.

\textbf{Project analysis:} The project’s first phase involves the construction of a 320,000ton zero-carbon ammonia device and a wind and photovoltaic power station of 1.35GW, with an investment of approximately ¥9 billion. The project’s first phase has an investment of ¥2.01 billion, mainly building 375MW of wind power and 56MW of energy storage, delivering the rated capacity for a maximum of four hours. Also, a new 220kV booster station will be built, and a green power line of about 70km from Wengqi to Yuanbaoshan will be built. The second phase plans to invest ¥4 billion, covering an area of 202 acres, with a designed annual output of 300,000 tons of synthetic ammonia and 10,000 tons of liquid ammonia. It is planned to purchase one hundred and twenty sets of water electrolysis hydrogen production equipment, start construction in August 2024, and put it into production in November 2026. The third phase plans to invest ¥9 billion, covering an area of 517 acres, with a designed annual output of 900,000 tons of synthetic ammonia and 10,000 tons of liquid hydrogen. It will purchase three hundred sets of electrolytic water hydrogen production equipment \cite{unknowp-j}. It is planned to start its construction in December 2026 and put it into production in June 2029 \cite{unknowp-k}.

\revpara{Table \ref{tab: Summary of demonstration projects for hydrogen production across the world} has been reconstructed to include project-status classification, source-year information, and a clearer distinction between reported metrics and annual hydrogen-output data.}

\begingroup
\tiny
\setlength{\tabcolsep}{1pt}
\renewcommand{\arraystretch}{1.15}
\setlength{\LTleft}{0pt}
\setlength{\LTright}{0pt}
\setlength{\LTcapwidth}{\textwidth}
\sloppy

\begin{longtable}{P{0.030\textwidth} P{0.105\textwidth} P{0.060\textwidth} P{0.125\textwidth} P{0.145\textwidth} P{0.155\textwidth} P{0.150\textwidth} P{0.155\textwidth}}
\caption{Summary of demonstration projects for hydrogen production across the world}
\label{tab: Summary of demonstration projects for hydrogen production across the world}\\

\toprule
\rev{No.} &
\rev{Project} &
\rev{Country / region} &
\rev{Project status / data basis} &
\rev{Main technology} &
\rev{Reported capacity / output} &
\rev{Annual H$_2$ output / normalization note} &
\rev{Main application / key outcome} \\
\midrule
\endfirsthead

\multicolumn{8}{c}{\tablename\ \thetable{} -- continued from previous page} \\
\toprule
\rev{No.} &
\rev{Project} &
\rev{Country / region} &
\rev{Project status / data basis} &
\rev{Main technology} &
\rev{Reported capacity / output} &
\rev{Annual H$_2$ output / normalization note} &
\rev{Main application / key outcome} \\
\midrule
\endhead

\midrule
\multicolumn{8}{r}{Continued on next page} \\
\endfoot

\bottomrule
\endlastfoot

\rev{1} &
\rev{HyNet} &
\rev{UK} &
\rev{Under development; project documents mainly 2021--2024; design data rather than verified operation} &
\rev{ATR + CCUS; hydrogen pipeline; salt-cavern hydrogen storage} &
\rev{Two hydrogen production units with total design capacity of 890 MW$_{\mathrm{th}}$; salt-cavern storage capacity: 35,000 t H$_2$} &
\rev{Not disclosed; storage capacity is not treated as annual production} &
\rev{Low-carbon hydrogen supply for industrial users, power plants, gas networks, refueling stations, and regional decarbonization} \\

\rev{2} &
\rev{H2Teesside} &
\rev{UK} &
\rev{Planned project in the original manuscript; later public status update indicates application withdrawal in Dec. 2025; announced/design data} &
\rev{Blue hydrogen from natural gas + CCUS} &
\rev{1.2 GW$_{\mathrm{th}}$ low-carbon hydrogen production facility; project document reports approx. 160,000 t low-carbon H$_2$/year} &
\rev{Approx. 160,000 t H$_2$/year as announced project output} &
\rev{Industrial decarbonization, fuel switching, heavy transport, and synthetic-fuel feedstock} \\

\rev{3} &
\rev{H2FUTURE} &
\rev{Austria} &
\rev{Completed demonstration, 2017--2021; project-level operational demonstration data} &
\rev{6 MW PEM electrolyzer; hydrogen-based metallurgy; grid services} &
\rev{6 MW PEM electrolyzer; 1,200 m$^3$ green H$_2$/h} &
\rev{Annual output not disclosed; hourly capacity is retained and not converted without utilization hours} &
\rev{Green hydrogen production for steel-sector decarbonization and grid-balancing services} \\

\rev{4} &
\rev{Demo4Grid} &
\rev{Austria} &
\rev{Completed / demonstration project, 2017--2023; demonstration and design-capacity data} &
\rev{Pressurized alkaline electrolyzer; grid-balancing services} &
\rev{4 MW pressurized alkaline electrolyzer; manuscript table reports 1,300 kg H$_2$/day} &
\rev{Annual output not disclosed; daily production rate is retained as reported} &
\rev{Green hydrogen production combined with grid frequency-control and balancing services} \\

\rev{5} &
\rev{Stegra Boden plant (formerly H2 Green Steel)} &
\rev{Sweden} &
\rev{Under construction; updated name and official status; production planned for 2026} &
\rev{Large-scale electrolysis; hydrogen-based direct reduction; green steel production} &
\rev{More than 700 MW electrolyzer capacity; initial green steel capacity: 2.5 Mt/year; planned ramp-up to around 5 Mt/year by 2030} &
\rev{Pure annual H$_2$ output not disclosed; steel-output target is reported separately} &
\rev{Green hydrogen used for green iron and green steel production, replacing coal-based reduction} \\

\rev{6} &
\rev{Utah Hydrogen Power Project / ACES Delta} &
\rev{USA} &
\rev{Under development; sources mainly 2023--2024; announced design and storage data} &
\rev{220 MW alkaline electrolysis; salt-cavern H$_2$ storage; hydrogen-capable gas turbines} &
\rev{Over 220 MW renewable energy conversion; up to 100 metric t H$_2$/day; more than 300 GWh clean energy storage; 840 MW hydrogen-capable power plant} &
\rev{Annual output not disclosed; maximum daily production and storage data are retained as reported} &
\rev{Large-scale seasonal hydrogen storage and dispatchable hydrogen-based power generation} \\

\rev{7} &
\rev{H2Hubs} &
\rev{USA} &
\rev{Policy / regional hub program selected in 2023; not treated as a single production project} &
\rev{Seven regional clean hydrogen hubs connecting producers, consumers, storage, delivery, and end-use infrastructure} &
\rev{Seven hubs; expected combined production of approximately 3 Mt clean H$_2$/year} &
\rev{Approx. 3 Mt H$_2$/year is a combined program-level target, not verified output from one project} &
\rev{National clean hydrogen supply-chain development for industry, transport, power, and storage} \\

\rev{8} &
\rev{EverWind Fuels} &
\rev{Canada} &
\rev{Planned / early development; sources mainly 2023--2025; announced output data} &
\rev{Renewable-powered electrolysis; green ammonia synthesis; port-based export} &
\rev{Official current source reports 240,000 t green ammonia/year for the Point Tupper project} &
\rev{Pure annual H$_2$ output not disclosed in current official project source; ammonia output is reported instead} &
\rev{Green ammonia production for export and industrial decarbonization; hydrogen is used as intermediate feedstock} \\

\rev{9} &
\rev{Asian / Australian Renewable Energy Hub, AREH} &
\rev{Australia} &
\rev{Planned / prospective project; BP exit reported in 2025; design and target data only} &
\rev{26 GW wind and solar power; large-scale electrolysis; green ammonia export} &
\rev{Up to 26 GW renewable power; at full capacity, around 1.6 Mt green H$_2$/year or 9 Mt green ammonia/year} &
\rev{Up to 1.6 Mt H$_2$/year is an announced full-scale target if configured for hydrogen production} &
\rev{Large-scale renewable hydrogen and ammonia export hub} \\

\rev{10} &
\rev{FH2R} &
\rev{Japan} &
\rev{Operational demonstration, 2020--present; demonstration and source-reported data} &
\rev{20 MW PV; 10 MW-class alkaline electrolyzer; hydrogen energy management system} &
\rev{20 MW solar power; 10 MW-class hydrogen production unit; up to 1,200 Nm$^3$ H$_2$/h} &
\rev{200 t H$_2$/year reported by H2Valleys; other sources report hourly capacity only} &
\rev{Solar-to-hydrogen production, storage, transport, demand response, and EMS demonstration} \\

\rev{11} &
\rev{Ordos smart zero-carbon green hydrogen project} &
\rev{China} &
\rev{Demonstration / under development; sources mainly 2022--2023; project and product-source data} &
\rev{Off-grid photovoltaic hydrogen production; PEM ``Hydrogen Boat'' system; digital monitoring platform} &
\rev{Manuscript reports 1,000 Nm$^3$ H$_2$/h; public company source confirms the Ordos zero-carbon smart hydrogen demonstration case and Hydrogen Boat digital platform} &
\rev{Annual output not disclosed; hourly capacity is retained as reported} &
\rev{Demonstration of off-grid photovoltaic hydrogen production and intelligent PEM electrolyzer operation} \\

\rev{12} &
\rev{Ningxia Ningdong renewable hydrogen project} &
\rev{China} &
\rev{Commercial demonstration / under development; sources mainly 2023--2024; project-reported data} &
\rev{620 MW-class PV; alkaline electrolysis; hydrogen refueling; coal-chemical integration} &
\rev{Two hydrogen plants with total capacity of 20,000 Nm$^3$/h; two refueling stations with total capacity of 4,000 kg/day} &
\rev{Approx. 4,500 t H$_2$/year reported in manuscript/project materials} &
\rev{Green hydrogen substitution in coal-chemical production and hydrogen heavy-truck refueling} \\

\rev{13} &
\rev{Chifeng wind-solar green hydrogen and ammonia project} &
\rev{China} &
\rev{Phased project; first phase operational in 2025; planned full scale around 2028; product-output target data} &
\rev{Wind- and solar-powered electrolysis; green ammonia synthesis; zero-carbon industrial gas production} &
\rev{Phase I: 320,000 t/year green ammonia; total planned capacity: approx. 1.52 Mt/year green hydrogen-ammonia / zero-carbon industrial gas products} &
\rev{Pure annual H$_2$ output not disclosed; product output is mainly reported as green ammonia / hydrogen-ammonia products} &
\rev{Large-scale green ammonia and zero-carbon industrial gas production for chemical, shipping, steel, and other sectors} \\

\multicolumn{8}{>{\columncolor{blue!18}}P{0.98\textwidth}}{\textit{Note:} Project status is classified as operational/completed demonstration, under construction, planned/announced project, or policy/regional hub program. Reported values are retained in their original units to avoid hidden assumptions. Annual H$_2$ output is provided only when explicitly reported by project sources or project documents. Hourly capacity, daily production rate, storage capacity, ammonia output, steel output, and regional hub targets are not treated as verified annual production unless the source reports such values directly.} \\

\end{longtable}
\endgroup

Table \ref{tab: Summary of demonstration projects for hydrogen production across the world} summarizes the current demonstration projects for hydrogen production, as mentioned above. \revpara{The revised table separates operational demonstrations, under-construction projects, planned or announced projects, and policy/regional hub programs. It also distinguishes verified annual hydrogen-output data from design capacities, announced targets, ammonia or steel product outputs, and storage-capacity indicators. Based on the preceding discussion, the reviewed cases can be grouped into three main deployment directions: (1) large-scale renewable or low-carbon hydrogen production and export-oriented hydrogen/ammonia hubs; (2) industrial decarbonization through hydrogen metallurgy, green ammonia, chemical production, and natural-gas substitution; and (3) grid-interactive and storage-based applications that improve the flexibility of renewable-integrated energy systems. This classification makes the cross-project comparison more transparent and avoids treating all cases as the same type of hydrogen-production project.}

In industrial production, the utilization of hydrogen for industrial gas preparation and metal smelting will significantly reduce environmental pollution and improve preparation efficiency \cite{tang2020a}. Hydrogen refueling stations are also gradually becoming popular in transportation \cite{greene2020a}. In the field of daily life, natural gas mixed with hydrogen and hydrogen energy connected to the grid can better meet the daily needs of residents while reducing emissions \cite{saeedmanesh2018a}. In addition, in terms of hydrogen energy storage, the current mainstream methods are compression storage and hydrogen storage tanks \cite{abe2019a}. With the increasing use of hydrogen energy in the future, achieving larger-scale, more efficient, and lower-cost hydrogen storage technology will also become the key to technological development. In short, from a long-term perspective, hydrogen energy still has enormous prospects and markets. In the future, in addition to continuing to explore and upgrade existing research, more research in hydrogen production technology innovation, hydrogen energy management should be conducted, hydrogen integration into multi-vector energy systems, and other directions. Through these means, establish a more complete, controllable, safer hydrogen energy industry chain to promote energy transformation and global climate targets.

\section{Conclusion}\label{sec: Conclusion}

Hydrogen is an essential means to facilitate energy transition in various sectors, holding significant market potential in the future. Introducing hydrogen into transportation and industrial manufacturing sectors can effectively reduce pollution, achieve decarbonization, and enhance efficiency, warranting further development of hydrogen policy and projects. \revpara{By 2022, 26 countries and economies had developed national hydrogen strategies, indicating that hydrogen had become an important component of energy-transition planning worldwide \cite{IEA_GlobalHydrogenReview2022,liu2022a}.} However, more comprehensive research on field demonstrations and industrial projects is needed to evaluate their long-term viability, technical performance, and economic impacts under real-world conditions.

This work has reviewed the current status of hydrogen production, storage, and application, along with an introduction, analysis, and summary of typical cases of global hydrogen demonstration projects. Research into demonstration and application projects are mainly in industrial manufacturing (e.g., hydrogen metallurgy and industrial gas production), transportation (e.g., hydrogen fuel cells and refueling stations), electricity grid applications (ensuring grid frequency’s static and dynamic security), and blending of hydrogen into natural gas systems. These applications highlight the versatility and potential of hydrogen energy in supporting sustainable development and reducing greenhouse gas emissions. {Additionally, these projects showcase the adaptability of hydrogen technologies across diverse geographic and industrial contexts, emphasizing the importance of tailored solutions that address specific regional energy demands and resource availability.}

This review clearly demonstrates there is room for improvement in efficiency, cost reduction, and scaling. This necessitates more profound research into technologies such as the design of advanced electrolyzers, the development of higher safety standards for hydrogen storage equipment, and the optimization of hydrogen distribution networks. Moreover, the integration of hydrogen with renewable energy sources, such as wind and solar power, presents significant opportunities to enhance energy system resilience and flexibility. Exploring hybrid systems that combine hydrogen with other energy carriers can further optimize energy management and reduce overall system costs. Additionally, enhancing the integration of hydrogen with renewable energy sources and exploring new hydrogen-based technologies will be critical for future advancements. Furthermore, on-site conditions for hydrogen projects are equally important to investigate, as they provide vital evidence of the application of hydrogen-integrated energy systems and offer insights into operational challenges and performance metrics under real-world conditions.

\revpara{Despite the promising progress demonstrated by these projects, several limitations and challenges still constrain the large-scale deployment of hydrogen technologies. From an economic perspective, high capital costs for electrolyzers, storage facilities, pipelines, refueling stations, and carbon capture infrastructure remain major barriers, especially when project revenues depend on uncertain electricity prices, low utilization factors, and immature hydrogen markets. Regulatory uncertainty is another important challenge, including inconsistent certification standards for low-carbon hydrogen, different safety requirements for hydrogen storage and blending, lengthy permitting procedures, and unclear market rules for cross-sector services such as grid balancing and industrial decarbonization. In addition, supply-chain constraints may slow global deployment, particularly in relation to critical materials for electrolyzers and fuel cells, manufacturing capacity, skilled labor, renewable electricity availability, water resources, and the development of ports, pipelines, salt caverns, and refueling networks. Addressing these challenges requires coordinated policy support, standardized reporting of project performance, bankable offtake mechanisms, and integrated planning across electricity, gas, industrial, and transport sectors.}

Future research could include more granular data on performance metrics and cost assessments to provide a clearer perspective on the scalability and economic viability of hydrogen in various applications. It is also recommended to focus on life-cycle assessments and environmental impact analyses to fully understand the sustainability of hydrogen projects. Additionally, interdisciplinary collaboration across sectors and international cooperation will play a pivotal role in accelerating the global adoption of hydrogen energy, promoting technological innovation, and achieving climate goals. The establishment of standardized methodologies for evaluating hydrogen projects can further enhance data comparability and support policy development. By addressing these areas, the hydrogen energy industry can establish a more complete, controllable, and safer industrial chain, driving the global transition toward a low-carbon future. Ultimately, hydrogen energy holds the promise of transforming the global energy landscape, fostering economic growth, and contributing significantly to the achievement of global climate targets.

\appendix

\vspace{-0.5cm}
 \bibliographystyle{elsarticle-num} 
 \renewcommand{\bibfont}{\footnotesize
}
 \bibliography{cas-refs}

@article{olabi2021a,
  citation-number = {1},
  author = {Olabi, A. and Abdelghafar, A.A. and Baroutaji, A. and Sayed, E.T. and Alami, A.H. and Rezk, H.},
  title = {Large-vscale hydrogen production and storage technologies: Current status and future directions},
  volume = {46},
  year= {2021},
  pages = {23498–23528},
  language = {en},
  journal = {International Journal of Hydrogen Energy}
}

@article{kova2021a,
  citation-number = {2},
  author = {Kovač, A. and Paranos, M. and Marciuš, D.},
  title = {Hydrogen in energy transition: A review},
  volume = {46},
  year = {2021},
  pages = {10016–10035},
  language = {en},
  journal = {International Journal of Hydrogen Energy}
}

@misc{IEA_GlobalHydrogenReview2022,
  author = {{International Energy Agency}},
  title = {Global Hydrogen Review 2022},
  url = {https://www.iea.org/reports/global-hydrogen-review-2022},
  year = {2022},
  language = {en}
}

@article{liu2015a,
  citation-number = {4},
  author = {Liu, H.-C. and Li, F.},
  title = {Novel high step-up {DC–DC} converter with an active coupled-inductor network for a sustainable energy system},
  volume = {30},
  year = {2015},
  pages = {6476–6482},
  language = {en},
  journal = {IEEE transactions on power electronics}
}

@article{ji2021a,
  citation-number = {6},
  author = {Ji, M. and Wang, J.},
  title = {Review and comparison of various hydrogen production methods based on costs and life cycle impact assessment indicators},
  volume = {46},
  year = {2021},
  pages = {38612–38635},
  language = {en},
  journal = {International Journal of Hydrogen Energy}
}

@article{cardinale2023a,
  citation-number = {7},
  author = {Cardinale, R.},
  title = {From natural gas to green hydrogen: Developing and repurposing transnational energy infrastructure connecting North Africa to Europe},
  volume = {181},
  year = {2023},
  language = {en},
  journal = {Energy Policy},
  number = {113623}
}

@article{thomas2020a,
  citation-number = {8},
  author = {Thomas, J.M. and Edwards, P.P. and Dobson, P.J. and Owen, G.P.},
  title = {Decarbonising energy: The developing international activity in hydrogen technologies and fuel cells},
  volume = {51},
  year = {2020},
  pages = {405–415},
  language = {en},
  journal = {Journal of Energy Chemistry}
}

@article{UKhydrogenstrategy,
  citation-number = {9},
  author = {UK},
  title = {{UK} Hydrogen Strategy},
  year = {2021},
  url = {https://assets.publishing.service.gov.uk/media/64c7e8bad8b1a70011b05e38/UK-Hydrogen-Strategy_web.pdf.},
  language = {en},
  journal = {HM Government}
}

@misc{policypaperUKhydrogenstrategy,
  citation-number = {10},
  title = {Policy paper UK hydrogen strategy},
  url = {https://www.gov.uk/government/publications/uk-hydrogen-strategy},
  year = {2024},
  language = {en}
}

@article{mclaughlin2023a,
  citation-number = {11},
  author = {McLaughlin, H. and Littlefield, A.A. and Menefee, M. and Kinzer, A. and Hull, T. and Sovacool, B.K.},
  title = {Carbon capture utilization and storage in review: Sociotechnical implications for a carbon reliant world},
  volume = {177},
  year = {2023},
  language = {en},
  journal = {Renewable and sustainable energy reviews},
  number = {113215}
}

@misc{newhubs,
  citation-number = {12},
  title = {New hubs will lead innovation towards a net zero energy sector},
  url = {https://www.ukri.org/news/new-hubs-will-lead-innovation-towards-a-net-zero-energy-sector/},
  year = {2024},
  language = {en}
}

@misc{roadmap2002a,
  citation-number = {13},
  author = {{U.S. Department of Energy}},
  title = {U.S. National Clean Hydrogen Strategy and Roadmap},
  url = {https://www.hydrogen.energy.gov/library/roadmaps-vision/clean-hydrogen-strategy-roadmap},
  year = {2023},
  language = {en}
}

@article{hassan2024a,
  citation-number = {14},
  author = {Hassan, Q. and Algburi, S. and Sameen, A.Z. and Salman, H.M. and Jaszczur, M.},
  title = {Green hydrogen: A pathway to a sustainable energy future},
  volume = {50},
  year = {2024},
  pages = {310–333},
  language = {en},
  journal = {International Journal of Hydrogen Energy}
}

@misc{liu2022a,
  citation-number = {15},
  author = {{Ministry of Economy, Trade and Industry}},
  title = {Basic Hydrogen Strategy},
  url = {https://www.meti.go.jp/shingikai/enecho/shoene_shinene/suiso_seisaku/pdf/20230606_5.pdf},
  year = {2023},
  language = {ja}
}

@article{lee2014a,
  citation-number = {16},
  author = {Lee, D.-H.},
  title = {Development and environmental impact of hydrogen supply chain in Japan: Assessment by the CGE-LCA method in Japan with a discussion of the importance of biohydrogen},
  volume = {39},
  year = {2014},
  pages = {19294–19310},
  language = {en},
  journal = {International Journal of Hydrogen Energy}
}

@article{tian2022a,
  citation-number = {17},
  author = {Tian, Z. and Lv, H. and Zhou, W. and Zhang, C. and He, P.},
  title = {Review on equipment configuration and operation process optimization of hydrogen refueling station},
  volume = {47},
  year = {2022},
  pages = {3033–3053},
  language = {en},
  journal = {International Journal of Hydrogen Energy}
}

@misc{huang2024a,
  citation-number = {18},
  author = {{National Development and Reform Commission}},
  title = {Medium- and Long-term Plan for Hydrogen Energy Industry Development (2021--2035)},
  url = {https://www.ndrc.gov.cn/xxgk/zcfb/ghwb/202203/t20220323_1320038.html},
  year = {2022},
  language = {zh}
}

@article{li2020a,
  citation-number = {19},
  author = {Li, Z. and Zhang, W. and Zhang, R. and Sun, H.},
  title = {Development of renewable energy multi-energy complementary hydrogen energy system (A Case Study in China): A review},
  volume = {38},
  year = {2020},
  pages = {2099–2127},
  language = {en},
  journal = {Energy Exploration \& Exploitation}
}

@article{shao2023a,
  citation-number = {20},
  author = {Shao, Z. and Cao, X. and Zhai, Q. and Guan, X.},
  title = {Risk-constrained planning of rural-area hydrogen-based microgrid considering multiscale and multi-energy storage systems},
  volume = {334},
  year = {2023},
  language = {en},
  journal = {Applied Energy},
  number = {120682}
}

@article{van2023a,
  citation-number = {21},
  author = {Van, L.P. and Chi, K.D. and Duc, T.N.},
  title = {Review of hydrogen technologies based microgrid: Energy management systems, challenges and future recommendations},
  volume = {48},
  year = {2023},
  pages = {14127–14148},
  language = {en},
  journal = {International Journal of Hydrogen Energy}
}

@article{fonseca2019a,
  citation-number = {22},
  author = {Fonseca, J.D. and Camargo, M. and Commenge, J.-M. and Falk, L. and Gil, I.D.},
  title = {Trends in design of distributed energy systems using hydrogen as energy vector: A systematic literature review},
  volume = {44},
  year = {2019},
  pages = {9486–9504},
  language = {en},
  journal = {International Journal of Hydrogen Energy}
}

@article{maestre2021a,
  citation-number = {23},
  author = {Maestre, V.M. and Ortiz, A. and Ortiz, I.},
  title = {Challenges and prospects of renewable hydrogen-based strategies for full decarbonization of stationary power applications},
  volume = {152},
  year = {2021},
  language = {en},
  journal = {Renewable and Sustainable Energy Reviews},
  number = {111628}
}

@article{yue2021a,
  citation-number = {24},
  author = {Yue, M. and Lambert, H. and Pahon, E. and Roche, R. and Jemei, S. and Hissel, D.},
  title = {Hydrogen energy systems: A critical review of technologies, applications, trends and challenges},
  volume = {146},
  year = {2021},
  language = {en},
  journal = {Renewable and sustainable energy reviews},
  number = {111180}
}

@article{martin2020a,
  citation-number = {25},
  author = {Martin, A. and Agnoletti, M.-F. and Brangier, E.},
  title = {Users in the design of Hydrogen Energy Systems: A systematic review},
  volume = {45},
  year = {2020},
  pages = {11889–11900},
  language = {en},
  journal = {International Journal of Hydrogen Energy}
}

@article{zhang2023a,
  citation-number = {26},
  author = {Zhang, S. and Wei, Y. and Guo, X. and Li, Z. and Song, X. and Blaabjerg, F.},
  title = {Overview of US patents for energy management of renewable energy systems with hydrogen},
  volume = {48},
  year = {2023},
  pages = {9574–9591},
  language = {en},
  journal = {International Journal of Hydrogen Energy}
}

@article{aunedi2022a,
  citation-number = {27},
  author = {Aunedi, M. and Yliruka, M. and Dehghan, S. and Pantaleo, A.M. and Shah, N. and Strbac, G.},
  title = {Multi-model assessment of heat decarbonisation options in the UK using electricity and hydrogen},
  volume = {194},
  year = {2022},
  pages = {1261–1276},
  language = {en},
  journal = {Renewable Energy}
}

@article{hermesmann2022a,
  citation-number = {28},
  author = {Hermesmann, M. and Green, Müller T.E. and Turquoise},
  title = {Blue, or Grey? Environmentally friendly Hydrogen Production in Transforming Energy Systems},
  volume = {90},
  year = {2022},
  language = {en},
  journal = {Progress in Energy and Combustion Science},
  number = {100996}
}

@article{incer-valverde2023a,
  citation-number = {29},
  author = {Incer-Valverde, J. and Korayem, A. and Tsatsaronis, G. and Morosuk, T.},
  title = {Colors” of hydrogen: Definitions and carbon intensity},
  volume = {291},
  year = {2023},
  language = {en},
  journal = {Energy Conversion and Management},
  number = {117294}
}

@article{huang2023a,
  citation-number = {30},
  author = {Huang, J. and Balcombe, P. and Feng, Z.},
  title = {Technical and economic analysis of different colours of producing hydrogen in China},
  volume = {337},
  year = {2023},
  language = {en},
  journal = {Fuel},
  number = {127227}
}

@article{hermesmann2021a,
  citation-number = {31},
  author = {Hermesmann, M. and Grübel, K. and Scherotzki, L. and Müller, T.},
  title = {Promising pathways: The geographic and energetic potential of power-to-x technologies based on regeneratively obtained hydrogen},
  volume = {138},
  year = {2021},
  language = {en},
  journal = {Renewable and sustainable energy reviews},
  number = {110644}
}

@article{kumar2022a,
  citation-number = {32},
  author = {Kumar, S.S. and Lim, H.},
  title = {An overview of water electrolysis technologies for green hydrogen production},
  volume = {8},
  year = {2022},
  pages = {13793–13813},
  language = {en},
  journal = {Energy reports}
}

@article{zuettel2004a,
  citation-number = {34},
  author = {Züttel, A.},
  title = {Hydrogen storage methods},
  volume = {91},
  year = {2004},
  pages = {157–172},
  language = {de},
  journal = {Naturwissenschaften}
}

@article{zheng2012a,
  citation-number = {35},
  author = {Zheng, J. and Liu, X. and Xu, P. and Liu, P. and Zhao, Y. and Yang, J.},
  title = {Development of high pressure gaseous hydrogen storage technologies},
  volume = {37},
  year = {2012},
  pages = {1048–1057},
  language = {de},
  journal = {International Journal of Hydrogen Energy}
}

@article{wang2023a,
  citation-number = {36},
  author = {Wang, C. and Ma, Y. and Gong, D. and Chen, W.},
  title = {The inhibition effects of helium gas on liquid hydrogen droplets group combustion},
  volume = {48},
  year = {2023},
  pages = {5333–5346},
  language = {en},
  journal = {International Journal of Hydrogen Energy}
}

@incollection{lim2010a,
  citation-number = {37},
  author = {Lim, K.L. and Kazemian, H. and Yaakob, Z. and Daud, W.W.},
  title = {Solid‐state materials and methods for hydrogen storage: a critical review},
  publisher = {Chemical Engineering \& Technology: Industrial Chemistry‐Plant Equipment‐Process Engineering‐Biotechnology},
  volume = {33},
  year = {2010},
  pages = {213–226},
  language = {en},
  booktitle = {Chemical Engineering \& Technology: Industrial Chemistry‐Plant Equipment‐Process Engineering‐Biotechnology}
}

@article{aziz2021a,
  citation-number = {38},
  author = {Aziz, M.},
  title = {Liquid hydrogen: A review on liquefaction, storage, transportation, and safety},
  volume = {14},
  year = {2021},
  language = {en},
  journal = {Energies},
  number = {5917}
}

@article{hassan2021a,
  citation-number = {42},
  author = {Hassan, I. and Ramadan, H.S. and Saleh, M.A. and Hissel, D.},
  title = {Hydrogen storage technologies for stationary and mobile applications: Review, analysis and perspectives},
  volume = {149},
  year = {2021},
  language = {en},
  journal = {Renewable and sustainable energy reviews},
  number = {111311}
}

@article{falcone2021a,
  citation-number = {47},
  author = {Falcone, P.M. and Hiete, M. and Sapio, A.},
  title = {Hydrogen economy and sustainable development goals: Review and policy insights},
  volume = {31},
  year = {2021},
  language = {en},
  journal = {Current opinion in green and sustainable chemistry},
  number = {100506}
}

@article{zainal2024a,
  citation-number = {48},
  author = {Zainal, B.S. and Ker, P.J. and Mohamed, H. and Ong, H.C. and Fattah, I. and Rahman, S.A.},
  title = {Recent advancement and assessment of green hydrogen production technologies},
  volume = {189},
  year = {2024},
  language = {en},
  journal = {Renewable and sustainable energy reviews},
  number = {113941}
}

@book{kumar2024a,
  citation-number = {49},
  author = {Kumar, R. and Kumar, A. and Sharma, A.},
  title = {Overview of Hydrogen Production Methods from Solar Energy},
  publisher = {CRC Press},
  year = {2024},
  pages = {223–245},
  language = {en},
  address = {Thermal Energy}
}

@article{dincer2015a,
  citation-number = {50},
  author = {Dincer, I. and Acar, C.},
  title = {Review and evaluation of hydrogen production methods for better sustainability},
  volume = {40},
  year = {2015},
  pages = {11094–11111},
  language = {en},
  journal = {International Journal of Hydrogen Energy}
}

@article{staffell2019a,
  citation-number = {51},
  author = {Staffell, I. and Scamman, D. and Abad, A.V. and Balcombe, P. and Dodds, P.E. and Ekins, P.},
  title = {The role of hydrogen and fuel cells in the global energy system},
  volume = {12},
  year = {2019},
  pages = {463–491},
  language = {en},
  journal = {Energy \& Environmental Science}
}

@article{ozturk2021a,
  citation-number = {52},
  author = {Ozturk, M. and Dincer, I.},
  title = {A comprehensive review on power-to-gas with hydrogen options for cleaner applications},
  volume = {46},
  year = {2021},
  pages = {31511–31522},
  language = {en},
  journal = {International Journal of Hydrogen Energy}
}

@article{aminudin2023a,
  citation-number = {53},
  author = {Aminudin, M. and Kamarudin, S. and Lim, B. and Majilan, E. and Masdar, M. and Shaari, N.},
  title = {An overview: Current progress on hydrogen fuel cell vehicles},
  volume = {48},
  year = {2023},
  pages = {4371–4388},
  language = {en},
  journal = {International Journal of Hydrogen Energy}
}

@article{colbertaldo2018a,
  citation-number = {54},
  author = {Colbertaldo, P. and Guandalini, G. and Campanari, S.},
  title = {Modelling the integrated power and transport energy system: The role of power-to-gas and hydrogen in long-term scenarios for Italy},
  volume = {154},
  year = {2018},
  pages = {592–601},
  language = {en},
  journal = {Energy}
}

@article{pastore2022a,
  citation-number = {55},
  author = {Pastore, L.M. and Basso, G.L. and Ricciardi, G. and Santoli, L.},
  title = {Synergies between Power-to-Heat and Power-to-Gas in renewable energy communities},
  volume = {198},
  year = {2022},
  pages = {1383–1397},
  language = {en},
  journal = {Renewable Energy}
}

@article{aminudin2023b,
  citation-number = {56},
  author = {Aminudin, M.A. and Kamarudin, S.K. and Lim, B.H. and Majilan, E.H. and Masdar, M.S. and Shaari, N.},
  title = {An overview: Current progress on hydrogen fuel cell vehicles},
  volume = {48},
  year = {2023},
  pages = {4371–4388},
  language = {en},
  journal = {International Journal of Hydrogen Energy}
}

@article{norouzi2021a,
  citation-number = {57},
  author = {Norouzi, N.},
  title = {An overview on the renewable hydrogen generation market},
  volume = {6},
  year = {2021},
  pages = {67–94},
  language = {en},
  journal = {INTERNATIONAL JOURNAL OF ENERGY STUDIES}
}

@article{taylor2022a,
  citation-number = {58},
  author = {Taylor, P. and Abeysekera, M. and Bian, Y. and Ćetenović, D. and Deakin, M. and Ehsan, A.},
  title = {An interdisciplinary research perspective on the future of multi-vector energy networks},
  volume = {135},
  year = {2022},
  language = {en},
  journal = {International Journal of Electrical Power \& Energy Systems},
  number = {107492}
}

@misc{fso,
  citation-number = {59},
  title = {Becoming the Future System Operator (FSO},
  url = {https://www.nationalgrideso.com/what-we-do/becoming-future-system-operator-fso},
  year = {2024},
  language = {en}
}

@misc{acctrans,
  citation-number = {60},
  author = {},
  title = {Accelerating the transition to a flexible, low carbon energy system},
  year = {2022},
  url = {https://www.nationalgrideso.com/document/266156/download.},
  language = {en},
  journal = {nation grid ESO}
}

@article{dehghan2024a,
  citation-number = {61},
  author = {Dehghan, S. and Aunedi, M. and Ameli, H. and Strbac, G.},
  title = {Whole-system value of electrified district heating networks in decarbonising heat sector in the UK},
  volume = {229},
  year = {2024},
  language = {en},
  journal = {Electric Power Systems Research},
  number = {110144}
}

@article{kang2020a,
  citation-number = {62},
  author = {Kang, J.-N. and Wei, Y.-M. and Liu, L.-C. and Han, R. and Yu, B.-Y. and Wang, J.-W.},
  title = {Energy systems for climate change mitigation: A systematic review},
  volume = {263},
  year = {2020},
  language = {en},
  journal = {Applied Energy},
  number = {114602}
}

@article{mavromatidis2019a,
  citation-number = {63},
  author = {Mavromatidis, G. and Orehounig, K. and Bollinger, L.A. and Hohmann, M. and Marquant, J.F. and Miglani, S.},
  title = {Ten questions concerning modeling of distributed multi-energy systems},
  volume = {165},
  year = {2019},
  language = {en},
  journal = {Building and Environment},
  number = {106372}
}

@article{he2023a,
  citation-number = {64},
  author = {He, Z. and Liu, C. and Wang, Y. and Wang, X. and Man, Y.},
  title = {Optimal operation of wind-solar-thermal collaborative power system considering carbon trading and energy storage},
  volume = {352},
  year = {2023},
  language = {en},
  journal = {Applied Energy},
  number = {121993}
}

@article{liu2022b,
  citation-number = {65},
  author = {Liu, Y. and Shi, H. and Guo, L. and Xu, T. and Zhao, B. and Wang, C.},
  title = {Towards long-period operational reliability of independent microgrid: A risk-aware energy scheduling and stochastic optimization method},
  volume = {254},
  year = {2022},
  language = {en},
  journal = {Energy},
  number = {124291}
}

@article{vivas2018a,
  citation-number = {66},
  author = {Vivas, F. and Heras, A. and Segura, F. and Andújar, J.},
  title = {A review of energy management strategies for renewable hybrid energy systems with hydrogen backup},
  volume = {82},
  year = {2018},
  pages = {126–155},
  language = {en},
  journal = {Renewable and sustainable energy reviews}
}

@article{li2022a,
  citation-number = {67},
  author = {Li, J. and Zou, W. and Yang, Q. and Bao, H.},
  title = {Towards net-zero smart system: An power synergy management approach of hydrogen and battery hybrid system with hydrogen safety consideration},
  volume = {263},
  year = {2022},
  language = {en},
  journal = {Energy Conversion and Management},
  number = {115717}
}

@misc{HyNetproject,
  citation-number = {68},
  title = {HyNet project},
  year = {2020},
  url = {https://hynet.co.uk/wp-content/uploads/2020/10/HyNet_NW-Vision-Document-2020_FINAL.pdf.},
  language = {en},
  journal = {HyNet North West. Hynet project}
}

@misc{hynetplant,
  citation-number = {69},
  title = {HyNet Low Carbon Hydrogen Plant Department for Business},
  year = {2021},
  url = {https://assets.publishing.service.gov.uk/media/621369ddd3bf7f4f027607af/Phase_2_Report_-_Progressive_Energy_-_HyNet_Low_Carbon_Hydrogen__3_.pdf.},
  language = {en},
  journal = {Energy and Industrial Strategy}
}

@article{robert2021a,
  citation-number = {70},
  author = {Robert, Castek and S, Harkin},
  title = {Evidence review for hydrogen for heat in buildings},
  year = {2021},
  url = {https://era.ed.ac.uk/bitstream/handle/1842/37964/CXC%20-%20Evidence%20review%20for%20hydrogen%20heat%20in%20buildings%20August%202021.pdf?sequence=3\&isAllowed=y.},
  language = {nl},
  journal = {Castek, Robert , Harkin Stephen}
}

@misc{Liverpool-Manchester,
  citation-number = {71},
  title = {The Liverpool-Manchester Hydrogen Cluster: A Low Cost, Deliverable Project},
  url = {https://hynet.co.uk/wp-content/uploads/2018/05/Liverpool-Manchester-Hydrogen-Cluster-Technical-Report-Cadent.pdf.},
  year = {2017},
  language = {en},
  journal = {Progressive Energy Ltd}
}

@misc{westnorthhynet,
  citation-number = {72},
  author = {West, HyNet North},
  title = {From Vision to Reality},
  url = {http://www.hynet.co.uk},
  year = {2023},
  language = {en}
}

@misc{hyneteconomics,
  citation-number = {73},
  title = {Potential Economic Impacts of the HyNet North West Project. HyNet Project},
  url = {https://hynet.co.uk/wp-content/uploads/2021/06/economic-impacts-report-040518.pdf.},
  year = {2023},
  language = {en}
}

@misc{unknown2021c,
  citation-number = {74},
  title = {Project report on Hynet North West},
  year = {2021},
  url = {https://hynet.co.uk/wp-content/uploads/2021/06/14368_CADENT_PROJECT_REPORT_AMENDED_v22105.pdf.},
  language = {en},
  journal = {Hynet Project}
}

@article{zhao2019a,
  citation-number = {75},
  author = {Zhao, Y. and McDonell, V. and Samuelsen, S.},
  title = {Influence of hydrogen addition to pipeline natural gas on the combustion performance of a cooktop burner},
  volume = {44},
  year = {2019},
  pages = {12239–12253},
  language = {en},
  journal = {International Journal of Hydrogen Energy}
}

@article{faheem2021a,
  citation-number = {76},
  author = {Faheem, H.H. and Tanveer, H.U. and Abbas, S.Z. and Maqbool, F.},
  title = {Comparative study of conventional steam-methane-reforming (SMR) and auto-thermal-reforming (ATR) with their hybrid sorption enhanced (SE-SMR \& SE-ATR) and environmentally benign process models for the hydrogen production},
  volume = {297},
  year = {2021},
  language = {en},
  journal = {Fuel},
  number = {120769}
}

@article{wang2022a,
  citation-number = {79},
  author = {Wang, H. and Tong, Z. and Zhou, G. and Zhang, C. and Zhou, H. and Wang, Y.},
  title = {Research and demonstration on hydrogen compatibility of pipelines: a review of current status and challenges},
  volume = {47},
  year = {2022},
  pages = {28585–28604},
  language = {en},
  journal = {International Journal of Hydrogen Energy}
}

@article{pathak2023a,
  citation-number = {80},
  author = {Pathak, P.K. and Yadav, A.K. and Padmanaban, S.},
  title = {Transition toward emission-free energy systems by 2050: Potential role of hydrogen},
  volume = {48},
  year = {2023},
  pages = {9921–9927},
  language = {en},
  journal = {International Journal of Hydrogen Energy}
}

@misc{unknown2021d,
  citation-number = {81},
  title = {Climate action in Austria Latest state of play},
  year = {2021},
  url = {https://www.europarl.europa.eu/RegData/etudes/BRIE/2021/696186/EPRS_BRI(2021)696186_EN.pdf.},
  language = {en},
  journal = {European Parliament}
}

@misc{h-a,
  citation-number = {82},
  author = {H2FUTURE},
  title = {Hydrogen from electrolysis for low carbon steelmaking},
  url = {https://stefan.schleicher.at/wp-content/uploads/2019/01/H2FUTURE-2018-ProjectPresentation-RudolfZauner.pdf},
  year = {2024},
  language = {en}
}

@misc{unknown-g,
  citation-number = {83},
  title = {H2FUTURE the official website of the project},
  url = {https://www.h2future-project.eu/},
  year = {2023},
  language = {en}
}

@misc{h-b,
  citation-number = {84},
  author = {H2FUTURE},
  title = {H2FUTURE Regulatory issues report},
  url = {https://www.h2future-project.eu/images/Publications/D9-5_Recommendations-to-regularory-bodies.pdf.},
  year = {2023},
  language = {sn}
}

@misc{unknown-h,
  citation-number = {85},
  title = {Hydrogen Technologies - greentec steel},
  url = {https://www.voestalpine.com/greentecsteel/en/breakthrough-technologies/},
  year = {2024},
  language = {en}
}

@article{carmo2013a,
  citation-number = {86},
  author = {Carmo, M. and Fritz, D.L. and Mergel, J. and Stolten, D.},
  title = {A comprehensive review on PEM water electrolysis},
  volume = {38},
  year = {2013},
  pages = {4901–4934},
  language = {en},
  journal = {International Journal of Hydrogen Energy}
}

@misc{unknown-i,
  citation-number = {87},
  title = {Production of Green Hydrogen},
  url = {https://www.h2future-project.eu/technology},
  year = {2023},
  language = {en}
}

@article{zhang2022a,
  citation-number = {88},
  author = {Zhang, Y. and Yue, Q. and Chai, X. and Wang, Q. and Lu, Y. and Ji, W.},
  title = {Analysis of process parameters on energy utilization and environmental impact of hydrogen metallurgy},
  volume = {361},
  year = {2022},
  language = {en},
  journal = {Journal of cleaner production},
  number = {132289}
}

@article{tang2020a,
  citation-number = {89},
  author = {Tang, J. and M-s, Chu and Li, F. and Feng, C. and Z-g, Liu and Y-s, Zhou},
  title = {Development and progress on hydrogen metallurgy},
  volume = {27},
  year = {2020},
  pages = {713–723},
  language = {lb},
  journal = {International Journal of Minerals, Metallurgy and Materials}
}

@misc{unknown-j,
  citation-number = {90},
  title = {Demonstration of 4MW Pressurized Alkaline Electrolyser for Grid Balancing Services},
  url = {https://cordis.europa.eu/project/id/736351},
  year = {2024},
  language = {en}
}

@misc{unknown-k,
  citation-number = {91},
  title = {Demo4Grid the official website of the project},
  url = {https://www.demo4grid.eu/},
  year = {2023},
  language = {en}
}

@misc{unknown-l,
  citation-number = {92},
  title = {Demo4Grid groundbreaking ceremony took place on the property of MPREIS in Völs near Innsbruck},
  url = {https://www.fen-systems.com/the-groundbreaking-ceremony-for-the-demo4grid-project-took-place-on-march-12-2020/},
  year = {2020},
  language = {en}
}

@misc{duo-a,
  citation-number = {93},
  title = {Technology DUo.Fuel Cell Electric Vehicle-to-grid feasibility : A technical analysis of aggregated units offering frequenc.},
  url ={https://pure.tudelft.nl/ws/portalfiles/portal/53334562/Robledo2019_Chapter_FuelCellElectricVehicle_to_Gri.pdf.},
  year = {2019},
  language = {en}
}

@misc{unknown2017b,
  citation-number = {94},
  title = {Europe's largest Single-Stack-Electrolyser for regulating the electricity grid and producing green hydr ogen for the food retailer},
  year = {2017},
  url = {https://www.demo4grid.eu/wp-content/uploads/2017/06/17-06-07-Final-PR-Demo4Grid-Kick-Off.pdf.},
  language = {en},
  journal = {Demo4Grid}
}

@misc{unknown-m,
  citation-number = {95},
  title = {Improvements to Integrate High Pressure Alkaline Electrolysers for Electricity/H2 production from R enewable Energies to Balance the Grid},
  url = {https://hydrogeneurope.eu/project/demo4grid},
  year = {2023},
  language = {en}
}

@misc{unknown-n,
  citation-number = {96},
  title = {Grid Integrated Multi Megawatt High Pressure Alkaline Electrolysers for Energy Applications},
  url = {https://cordis.europa.eu/article/id/413231-green-hydrogen-promises-more-renewable-integration-with-more-stable-grids},
  year = {2023},
  language = {en}
}

@misc{unknown-o,
  citation-number = {97},
  title = {Intermountain Power Agency Orders MHPS JAC Gas Turbine Technology for Renewable-Hydrogen Energy Hub},
  url = {https://power.mhi.com/regions/amer/news/200310.html},
  year = {2024},
  language = {en}
}

@misc{unknown-p,
  citation-number = {98},
  title = {Intermountain Power Plant Renewal},
  url = {https://www.nsenergybusiness.com/projects/intermountain-power-plant-renewal/},
  year = {2024},
  language = {en}
}

@misc{storage-a,
  citation-number = {99},
  title = {Advanced Clean Energy Storage},
  url = {https://www.energy.gov/lpo/advanced-clean-energy-storage},
  year = {2022},
  language = {en}
}

@misc{hydrogen-a,
  citation-number = {100},
  author = {HYDROGEN, C.O.N.V.E.R.S.I.O.N.O.F.I.N.T.E.R.M.O.U.N.T.A.I.N.P.O.W.E.R.P.R.O.J.E.C.T.T.O.G.R.E.E.N.},
  url = {https://www.ghcoalition.org/green-hydrogen-at-scale},
  year = {2024},
  language = {en}
}

@misc{unknown-q,
  citation-number = {101},
  title = {{ACES} Delta's Giant Utah Salt Cavern Hydrogen Storage Project Gets \$504M Conditional DOE Loan Guarantee},
  url = {https://www.powermag.com/aces-deltas-giant-utah-salt-cavern-hydrogen-storage-project-gets-504m-conditional-doe-loan-guarantee/},
  year = {2024},
  language = {en}
}

@article{miocic2023a,
  citation-number = {102},
  author = {Miocic, J. and Heinemann, N. and Edlmann, K. and Scafidi, J. and Molaei, F. and Alcalde, J.},
  title = {Underground hydrogen storage: a review},
  volume = {528},
  year = {2023},
  pages = {73–86},
  language = {en},
  journal = {Geological Society, London, Special Publications}
}

@misc{renewed-a,
  citation-number = {105},
  author = {Renewed, I.P.P.},
  url = {https://www.ipautah.com/ipp-renewed/},
  year = {2024},
  language = {en}
}

@misc{unknown2024a,
  citation-number = {106},
  title = {Asian Renewable Energy Hub},
  url = {https://ammoniaenergy.org/organization/asian-renewable-energy-hub/},
  year = {2024},
  language = {en}
}

@misc{unknown-t,
  citation-number = {107},
  title = {Tier 1 projects in Australia and the Middle East},
  url = {https://intercontinentalenergy.com/portfolio-of-projects/#AREH},
  year = {2024},
  language = {en}
}

@misc{unknown-u,
  citation-number = {108},
  title = {Renewable energy hub in Australia},
  url = {https://www.bp.com/en_au/australia/home/media/press-releases/bp-to-lead-and-operate-one-of-the-worlds-largest-renewables-and-green-hydrogen-energy-hubs-based-in-western-australia.html},
  year = {2024},
  language = {en}
}

@misc{sciences-a,
  citation-number = {109},
  author = {NW Interconnected Power Pty Ltd},
  title = {Asian Renewable Energy Hub – Section 38 Referral Supporting Information},
  url= {https://www.epa.wa.gov.au/sites/default/files/Referral_Documentation/Supportiing%20Information.pdf},
  year = {2017},
  language = {en}
}

@misc{unknown-v,
  citation-number = {110},
  title = {bp to lead and operate one of the world's largest renewables and green hydrogen energy hubs based in western Australia},
  url = {https://www.bp.com/en/global/corporate/news-and-insights/press-releases/bp-to-lead-and-operate-one-of-the-worlds-largest-renewables-and-green-hydrogen-energy-hubs-based-in-western-australia.html},
  year = {2024},
  language = {en}
}

@misc{unknown-w,
  citation-number = {111},
  title = {Australian government to fast-track 26GW hydrogen + ammonia ‘Asian Renewable Energy Hub’},
  url = {https://www.energy-storage.news/australian-government-to-fast-track-26gw-hydrogen-ammonia-asian-renewable-energy-hub/},
  year = {2024},
  language = {en}
}

@misc{bp-a,
  citation-number = {112},
  author = {BP TO LEAD, Office and RENEWABLES, O.P.E.R.A.T.E.O.N.E.O.F.T.H.E.W.O.R.L.D.’S.L.A.R.G.E.S.T. and AUSTRALIA, G.R.E.E.N.H.Y.D.R.O.G.E.N.E.N.E.R.G.Y.H.U.B.S.B.A.S.E.D.I.N.W.E.S.T.E.R.N.},
  url = {https://www.bp.com/content/dam/bp/country-sites/en_au/australia/home/media/media-releases/bp-to-lead-and-operate-one-of-the-worlds-largest-renewables-and-green-hydrogen-energy-hubs-based-in-western-australia.pdf.},
  year = {2022}
}

@misc{unknown-x,
  citation-number = {113},
  title = {Asian Renewable Energy Hub (AREH},
  url = {https://www.nsenergybusiness.com/projects/asian-renewable-energy-hub-areh-pilbara/},
  year = {2024},
  language = {ms},
  journal = {Pilbara}
}

@article{yuezba2022a,
  citation-number = {114},
  author = {Yüzbaşıoğlu, A.E. and Avşar, C. and Gezerman, A.O.},
  title = {The current situation in the use of ammonia as a sustainable energy source and its industrial potential},
  volume = {5},
  year = {2022},
  language = {en},
  journal = {Current Research in Green and Sustainable Chemistry},
  number = {100307}
}

@misc{unknown-y,
  citation-number = {115},
  title = {Asian Renewable Energy Hub},
  url = {https://www.minister.industry.gov.au/ministers/karenandrews/media-releases/job-creating-energy-hub-given-major-status-backing?utm_source=chatgpt.com},
  year = {2020},
  language = {en}
}

@misc{unknown-z,
  citation-number = {116},
  title = {Energy data},
  url = {https://www.energy.gov.au/energy-data/australian-energy-statistics/electricity-generation},
  year = {2024},
  language = {en}
}

@misc{unknowo-a,
  citation-number = {117},
  title = {Japan has taken an important step towards a "hydrogen energy economy"},
  url = {https://iauto.ifeng.com/news/quanmeiti/20181008/1241868.shtml?\&back},
  year = {2024},
  language = {en}
}

@article{khan2022a,
  citation-number = {118},
  author = {Khan, U. and Yamamoto, T. and Sato, H.},
  title = {Understanding the discontinuance trend of hydrogen fuel cell vehicles in Japan},
  volume = {47},
  year = {2022},
  pages = {31949–31963},
  language = {en},
  journal = {International Journal of Hydrogen Energy}
}

@misc{unknowo-b,
  citation-number = {119},
  title = {The world's largest photovoltaic hydrogen production project},
  url = {http://en.hfch.cn/},
  year = {2024},
  language = {en}
}

@misc{unknowo-c,
  citation-number = {120},
  title = {Fukushima Hydrogen Energy Research Field(FH2R},
  url = {https://www.nedo.go.jp/english/index.html},
  year = {2024},
  language = {en}
}

@misc{unknowo-d,
  citation-number = {121},
  title = {Asahi Kasei starts construction of alkaline water electrolysis pilot test plant for hydrogen production},
  url = {https://www.asahi-kasei.com/news/2022/e221107.html},
  year = {2024},
  language = {en}
}

@misc{unknowo-e,
  citation-number = {122},
  title = {The world's largest-class hydrogen production, Fukushima Hydrogen Energy Research Field (FH2R) now is completed at Namie town in Fukushima},
  url = {https://www.nedo.go.jp/english/news/AA5en_100422.html},
  year = {2024},
  language = {en}
}

@misc{unknowo-f,
  citation-number = {123},
  title = {World's largest green hydrogen plant breaks ground in Ordos},
  url = {https://www.chinadaily.com.cn/a/202302/17/WS63ef4bcea31057c47ebaf741.html},
  year = {2024},
  language = {en}
}

@book{group2023a,
  citation-number = {124},
  author = {Group, T.B.C. and ttoAO, Minggao},
  title = {China hydrogen industry Outlook},
  publisher = {Boston Consulting Group},
  year = {2023},
  url = {https://web-assets.bcg.com/a4/e9/7bdf588340169730c3dad382b4b2/bcg-china-hydrogen-industry-outlook-cn-aug-2023.pdf},
  language = {cn}
}

@misc{unknown2023a,
  citation-number = {125},
  title = {BCG's outlook for China's hydrogen industry},
  year = {2023},
  url = {https://web-assets.bcg.com/d4/2a/657fc0544c4e85a6c7533aed18fa/bcgxouyang-minggao-china-hydrogen-industry-outlook-en.pdf},
  language = {en}
}

@misc{unknowo-g,
  citation-number = {126},
  title = {Haide Hydrogen Energy and Huichuan Technology reached a strategic cooperation to create a hydrogen energy industry demonstration},
  url = {https://mp.weixin.qq.com/s/2cHHKaL_odq9PATtUBVj1g},
  year = {2022},
  language = {cn}
}

@misc{unknowo-i,
  citation-number = {128},
  title = {3.98kWh/Nm3 ! Hyde Hydrogen Energy "Hydrogen Boat" refreshes the industry's hydrogen production performance to a new height},
  url = {https://www.escn.com.cn/20230709/576100690bb04af19c1a0bd395462f79/c.html},
  year = {2023},
  language = {en}
}

@misc{unknowo-j,
  citation-number = {129},
  title = {Hyde Hydrogen Energy released a new generation of water electrolysis hydrogen production system "Hydrogen Boat", writing a new chapter in the cost reduction and efficiency increase of green hydrogen manufacturing},
  url = {https://www.ljzfin.com/news/info/69860.html},
  year = {2024},
  language = {en}
}

@misc{unknowo-k,
  citation-number = {130},
  title = {Hyde Hydrogen Energy's new generation of water electrolysis hydrogen production system "Hydrogen Boat},
  url = {https://www.htech360.com/a/6723},
  year = {2024},
  language = {en}
}

@misc{unknowo-l,
  citation-number = {131},
  title = {Hyde Hydrogen Energy launched a standardized alkaline electrolysis hydrogen production system under pressure that disrupts the industry “Hydrogen boat"},
  url = {http://www.tanjiaoyi.com/article-51169-1.html},
  year = {2024},
  language = {en}
}

@misc{unknowo-n,
  citation-number = {133},
  title = {World’s largest green hydrogen project — China's 260MW Kuqa facility — to be commissioned at the end of May},
  url = {https://www.hydrogeninsight.com/production/world-s-largest-green-hydrogen-project-chinas-260mw-kuqa-facility-to-be-commissioned-at-the-end-of-may/2-1-1457242},
  year = {2024},
  language = {en}
}

@misc{unknowo-q,
  citation-number = {136},
  title = {Photovoltaic Systems of Ningdong Renewable Hydrogen and Carbon Emission Reduction Demonstration Project of CHN Energy Connected to Grid},
  url = {http://en.sasac.gov.cn/2023/07/03/c_15508.htm},
  year = {2024},
  language = {en}
}

@misc{unknowo-r,
  citation-number = {137},
  title = {CHN Energy’s Ningdong Renewable Hydrogen and Carbon Emission Reduction Project Connected to Photovoltaic Grid},
  url = {https://www.ceic.com/gjnyjtwwEn/xwzx/202307/23d916c8fca140d3ba6b2ce8d6281ac3.shtml},
  year = {2024},
  language = {en}
}

@misc{unknowo-s,
  citation-number = {138},
  title = {Ningdong Renewable Energy Hydrogen Production Demonstration Project},
  url = {https://xueqiu.com/8995599040/193714079},
  year = {2024},
  language = {en}
}

@misc{unknowo-t,
  citation-number = {139},
  title = {The Yongli Hydrogenation Station in China Energy Guohua Investment Ningdong Renewable Hydrogen Carbon Reduction Demonstration Zone Passed its Mid-term Acceptance Test},
  url = {https://www.ceic.com/gjnyjtwwEn/xwzx/202310/53a26ddbb2fb411bafe9af8920834635.shtml},
  year = {2024},
  language = {en}
}

@misc{unknowo-u,
  citation-number = {140},
  title = {National Energy Ningdong Renewable Hydrogen Carbon Emission Reduction Demonstration Zone},
  url = {http://news.sohu.com/a/707622501_120142535},
  year = {2023},
  language = {cn}
}

@article{shi2023a,
  citation-number = {141},
  author = {Shi, Yuxuan and Zhang, Zhiquan},
  title = {Domestic and foreign hydrogen industry development practice comparison and the enlightenment to our country},
  volume = {52},
  year = {2023},
  pages = {2377–2380},
  language = {en},
  journal = {Applied Chemical Industry}
}

@misc{unknowo-v,
  citation-number = {142},
  title = {140 million! Guohua invested in the Ningdong Green Hydrogen Demonstration Project and 21 hydrogen production equipment won the bid},
  url = {https://news.bjx.com.cn/html/20230427/1303871.shtml},
  year = {2024},
  language = {en}
}

@misc{unknowo-w,
  citation-number = {143},
  title = {EPC bidding for the Ningdong Renewable Hydrogen Carbon Emission Reduction Demonstration Zone Project},
  url = {https://news.bjx.com.cn/html/20230110/1282069.shtml},
  year = {2024},
  language = {en}
}

@misc{unknowo-x,
  citation-number = {144},
  title = {National Energy Group participates in the build! The hydrogen refueling station of the Ningdong Renewable Hydrogen Carbon Emission Reduction Demonstration Zone Project was completed by CCCC!},
  url = {https://h2.in-en.com/html/h2-2429059.shtml},
  year = {2024},
  language = {en}
}

@misc{unknowp-f,
  citation-number = {152},
  title = {The Hydrogen Stream: Malaysia, Singapore mulling hydrogen pipeline},
  url = {https://www.pv-magazine.com/2023/10/24/the-hydrogen-stream-malaysia-singapore-mulling-hydrogen-pipeline/},
  year = {2024},
  language = {en}
}

@misc{unknowp-g,
  citation-number = {153},
  title = {Global ammonia market size is 180 million tons per year. How much of that will be occupied by green ammonia?},
  url = {http://www.asiachem.org/en/fcv_20220915},
  year = {2024},
  language = {en}
}

@misc{unknowp-h,
  citation-number = {154},
  title = {Wind and solar power generation hydrogen production energy storage},
  url = {https://www.pengky.cn/hydrogen_energy/09-wind-solar-energy-storage/wind-solar-energy-storage.html},
  year = {2024},
  language = {en}
}

@misc{unknowp-i,
  citation-number = {155},
  title = {Envision began construction of its 40 billion Rmb green ammonia plant},
  url = {https://h2ce.ru/en/news/industry-news/envision-began-construction-of-its-40-billion-rmb-green-ammonia-plant},
  year = {2024},
  language = {en}
}

@misc{unknowp-j,
  citation-number = {156},
  title = {Two wind and photovoltaic base projects in Inner Mongolia started construction},
  url = {https://www.seetao.com/details/226280.html},
  year = {2024},
  language = {en}
}

@misc{unknowp-k,
  citation-number = {157},
  title = {The world's first zero-carbon ammonia project settled in Chifeng},
  url = {https://www.seetao.com/details/147657.html},
  year = {2024},
  language = {en}
}

@article{greene2020a,
  citation-number = {158},
  author = {Greene, D.L. and Ogden, J.M. and Lin, Z.},
  title = {Challenges in the designing, planning and deployment of hydrogen refueling infrastructure for fuel cell electric vehicles},
  volume = {6},
  year = {2020},
  language = {en},
  journal = {ETransportation},
  number = {100086}
}

@article{saeedmanesh2018a,
  citation-number = {159},
  author = {Saeedmanesh, A. and Mac Kinnon, M.A. and Brouwer, J.},
  title = {Hydrogen is essential for sustainability},
  volume = {12},
  year = {2018},
  pages = {166–181},
  language = {en},
  journal = {Current Opinion in Electrochemistry}
}

@article{abe2019a,
  citation-number = {160},
  author = {Abe, J.O. and Popoola, A. and Ajenifuja, E. and Popoola, O.M.},
  title = {Hydrogen energy, economy and storage: Review and recommendation},
  volume = {44},
  year = {2019},
  pages = {15072–15086},
  language = {en},
  journal = {International Journal of Hydrogen Energy}
}

@article{isaac2019a,
  citation-number = {161},
  author = {Isaac, Tommy.},
  title = {HyDeploy: The UK’s first hydrogen blending deployment project},
  year = {2019},
  pages = {114-125},
  language = {en},
  journal = {Clean Energy}
}

@article{andy2020a,
  citation-number = {162},
  author = {Brown, Andy. and Dave Parkin.},
  title = {HyNet: Demonstrating an Integrated Hydrogen Economy},
  year = {2020},
  language = {en},
  journal = {The Chemical Engineer}
}

@article{badejo2024a,
  citation-number = {163},
  author = {Badejo, Adeshina O., et al.},
  title = {Assessment of the Environmental Risks and Policies Associated with Carbon Capture and Storage in the United Kingdom–A Case Study of the HyNet North West Project},
  booktitle = {Abu Dhabi International Petroleum Exhibition and Conference},
  year = {2024},
  language = {en},
  journal = {SPE}
}

@article{joan2002a,
  citation-number = {164},
  author = {Joan M. Ogden.},
  title = {Hydrogen: The Fuel of the Future?},
  volume = {55},
  year = {2002},
  pages = {69–75},
  language = {en},
  journal = {Physics Today},
  number = {4}
}

@article{ogwumike2024a,
  citation-number = {165},
  author={Ogwumike, Chris and Akponeware, Anderson and Oyewole, Adepeju and Dawood, Huda and Pinedo-Cuenca, Ruben and Ling-Chin, Janie and Roskilly, Anthony Paul and Dawood, Nashwan},
  title = {Transitioning or tinkering at a net-zero economy? Introducing an assessment framework for industrial cluster decarbonisation in the United Kingdom},
  year = {2024},
  language = {en},
  journal = {Energy Research \& Social Science},
}

@article{herman2024a,
  citation-number = {166},
  author = {Herman, Kyle S. Benjamin K. Sovacool and Marfuga Iskandarova.},
  title = {Imagining a net-zero Teesside: actors, networks, and expectations in industrial decarbonisation megaprojects},
  year = {2024},
  language = {en},
  journal = {Environmental Research Communications}
}

@article{soyoung2024a,
  citation-number = {167},
  author = {Oh, Soyoung. and Mohammed Al-Juaied.},
  title = {Decarbonizing industrial hubs and clusters: Towards an integrated framework of green industrial policies},
  year = {2024},
  language = {en},
  journal = {Energy Research \& Social Science},
}

@misc{unknowp-l,
  citation-number = {168},
  url = {https://www.bp.com.cn/zh_cn/china/home/news/press-releases/news-03-18-2021.html},
  year = {2021},
  language = {cn}
}

@article{bade2024a,
  citation-number = {172},
  author = {Bade, Shree Om. and Olusegun Stanley Tomomewo.},
  title = {A review of governance strategies, policy measures, and regulatory framework for hydrogen energy in the United States},
  volume = {78},
  year = {2024},
  pages = {1363-1381},
  language = {en},
  journal = {International Journal of Hydrogen Energy},
  number = {9}
}

@misc{unknowp-m,
  citation-number = {173},
  title = {DOE Regional Clean Hydrogen Hubs Program (H2Hubs)},
  year = {2023},
  url = {https://www.brattle.com/wp-content/uploads/2023/12/DOE-Regional-Clean-Hydrogen-Hubs-Program-H2Hubs.pdf.},
  language = {en}
}

@article{brunik2024a,
  citation-number = {174},
  author = {Brunik, K., et al.},
  title = {Potential for large-scale deployment of offshore wind-to-hydrogen systems in the United States},
  volume = {2767},
  year = {2024},
  language = {en},
  journal = {Journal of Physics: Conference Series},
  number = {6}
}

@inproceedings{Elmfeldt2024,
  citation-number = {175},
  author = {T. Elmfeldt and Y. Arafat and L. B. Tjernberg and A. Lugnet and G. Nyström},
  title = {Sector - coupling Green Hydrogen to Electrify Steel Production – A Case Study at Ovako Hofors},
  booktitle = {2024 18th International Conference on Probabilistic Methods Applied to Power Systems (PMAPS)},
  year = {2024},
  pages = {1 - 6},
  address = {Auckland, New Zealand},
}

@misc{unknowp-n,
  citation-number = {176},
  title = {Decarbonizing at scal},
  year = {2020},
  url = {https://stegra.com/},
  language = {en}
}

@misc{unknowp-o,
  citation-number = {177},
  title = {Introduction to H2 Green Steel},
  year = {2021},
  url = {https://www.eib.org/en/projects/all/20200902},
  language = {en}
}

@article{Ledari2023,
    citation-number = {178},
    author = {Ledari, Masoumeh Bararzadeh and others},
    title = {Greening steel industry by hydrogen: Lessons learned for the developing world},
    journal = {International Journal of Hydrogen Energy},
    year = {2023}
}

@misc{unknowp-p,
  citation-number = {179},
  title = {Economic contribution study report},
  year = {2024},
  url = {https://everwindfuels.com/economic_contribution_study_report.pdf}
}

@article{Graney2023a,
    citation-number = {180},

    author = {Graney, E.},
    title = {Ottawa signals green hydrogen support with financial boost for Nova Scotia project},
    journal = {The Globe and Mail web edition},
    year = {2023},
    url = {https://www.theglobeandmail.com/business/article-ottawa-signals-green-hydrogen-support-with-financial-boost-for-nova},
    pages = {NA - NA}
}

@misc{VanPraet2023,
    citation-number = {181},
    author = {Deloitte},
    title = {Economic Impact Assessment of EverWind’s Point Tupper Green Hydrogen and Ammonia Project}, 
    url = {https://everwindfuels.com/economic_contribution_study_report.pdf},
    year = {2024}
   
}

@misc{unknowp-q,
  citation-number = {182},
  title = {Australia backs 26GW wind and solar-powered green hydrogen hub},
  url = {https://www.windpowermonthly.com/article/1698179/australia-backs-26gw-wind-solar-powered-green-hydrogen-hub},
  year = {2020},
  language = {en}
}

@misc{unknowp-r,
  citation-number = {183},
  title = {A total investment of 3.8 billion yuan Guoneng Ningdong Renewable Hydrogen and Carbon Emission Reduction Demonstration Zone Project Pre-approval Planning Announcement},
  url = {https://www.sohu.com/a/754162404_121123735},
  year = {2024},
  language = {cn}
}

@misc{unknowp-s,
  citation-number = {183},
  title = {Hydrogen Boat, Hydrogen Energy's new generation of electrolytic water hydrogen production systems, makes its debut},
  url = {https://mp.weixin.qq.com/s/MNujRYb9ErNH0GwrTybniw},
  year = {2022},
  language = {cn}
}

@misc{unknowp-t,
  citation-number = {184},
  title = {See how Hydrogen Boat™, the "Super Hydrogen System", is setting new heights for hydrogen production in the industry},
  url = {https://mp.weixin.qq.com/s/l3-0laBkh1pjE9sACIKFRQ},
  year = {2023},
  language = {cn}
}

@misc{unknowp-u,
  citation-number = {185},
  title = {Hydrogen ESG Annual Practice Summary},
  url = {https://mp.weixin.qq.com/s/zSO6SniJnUeBJ4FHhrM3OA},
  year = {2023},
  language = {cn}
}

@misc{unknowp-v,
  citation-number = {186},
  title = {Green ammonia at oil and gas scale: the 15 GW Asian Renewable Energy Hub},
  url = {https://ammoniaenergy.org/articles/green-ammonia-at-oil-and-gas-scale-the-15-gw-asian-renewable-energy-hub/},
  year = {2020},
  language = {en}
}

@misc{unknowp-w,
  citation-number = {187},
  title = {Australian Renewable Energy Hub},
  url = {https://research.csiro.au/hyresource/australian-renewable-energy-hub/},
  year = {2024},
  language = {en}
}

@article{chehade2019ptx,
  title   = {Review and analysis of demonstration projects on power-to-X pathways in the world},
  author  = {Chehade, Zaher and Mansilla, Christine and Lucchese, Paul and Hilliard, Samantha and Proost, Joris},
  journal = {International Journal of Hydrogen Energy},
  volume  = {44},
  number  = {51},
  pages   = {27637--27655},
  year    = {2019},
  doi     = {10.1016/j.ijhydene.2019.08.260}
}

@article{wulf2020ptx,
  title   = {Review of Power-to-X Demonstration Projects in Europe},
  author  = {Wulf, Christina and Zapp, Petra and Schreiber, Andrea},
  journal = {Frontiers in Energy Research},
  volume  = {8},
  pages   = {191},
  year    = {2020},
  doi     = {10.3389/fenrg.2020.00191}
}

@techreport{DOEWaterElectrolysisAssessment2024,
  title        = {Hydrogen Shot: Water Electrolysis Technology Assessment},
  author       = {{U.S. Department of Energy}},
  institution  = {U.S. Department of Energy},
  year         = {2024},
  url          = {https://www.energy.gov/sites/default/files/2024-12/hydrogen-shot-water-electrolysis-technology-assessment.pdf}
}

@misc{DOEFuelCellComparison,
  title        = {Comparison of Fuel Cell Technologies},
  author       = {{U.S. Department of Energy}},
  year         = {2025},
  note         = {Accessed 2026-05-24},
  url          = {https://www.energy.gov/cmei/fuels/comparison-fuel-cell-technologies}
}

@techreport{DOEMediumScaleCHP2011,
  title        = {Medium-scale CHP Fuel Cell System Targets},
  author       = {{U.S. Department of Energy}},
  institution  = {U.S. Department of Energy},
  year         = {2011},
  url          = {https://www.hydrogen.energy.gov/docs/hydrogenprogramlibraries/pdfs/11014_medium_scale_chp_target.pdf}
}

@misc{DOEHydrogenStorage,
  title = {Hydrogen Storage},
  author = {{U.S. Department of Energy}},
  url = {https://www.energy.gov/cmei/fuels/hydrogen-storage},
  year = {2024},
  note = {Accessed 2026}
}



\end{document}